\documentclass[journal]{IEEEtran}

\ifCLASSINFOpdf
\usepackage[pdftex]{graphicx}
\else
\fi

\usepackage{amsmath}
\usepackage{amssymb}
\usepackage{amsthm}
\usepackage{color}
\usepackage{cases}
\usepackage{empheq}
\usepackage{algorithmic}
\usepackage[linesnumbered]{algorithm2e}
\usepackage{array}

\DeclareMathOperator*{\argmin}{arg\!\min}
\usepackage{epsfig}
\usepackage{epstopdf}
\usepackage{subcaption}
\newcommand{\vertiii}[1]{{\left\vert\kern-0.25ex\left\vert\kern-0.25ex\left\vert #1
    \right\vert\kern-0.25ex\right\vert\kern-0.25ex\right\vert}}

\begin{document}

\title{A Reference Governor for Nonlinear Systems with Disturbance Inputs
Based on Logarithmic Norms and Quadratic Programming}

\author{Nan~Li,~\IEEEmembership{Member,~IEEE,}
        Ilya~Kolmanovsky,~\IEEEmembership{Fellow,~IEEE,}
        Anouck~Girard,~\IEEEmembership{Senior Member,~IEEE}% <-this % stops a space
\thanks{This research is supported by the National Science Foundation under Award Number CNS 1544844 to the University of Michigan and by NASA under Cooperative Agreement NNX16AH81A.}}

\markboth{}%
{Li \MakeLowercase{\textit{et al.}}}

\maketitle

\begin{abstract}
This note describes a reference governor design for a continuous-time
nonlinear system with an additive disturbance. The design is based on predicting the response of the nonlinear system by the response of a linear model with a set-bounded prediction error, where a state-and-input dependent bound on the prediction error is explicitly characterized using logarithmic norms. The online optimization is reduced to a convex quadratic program with linear inequality constraints. Two numerical examples are reported.
\end{abstract}

\begin{IEEEkeywords}
Reference governors; Nonlinear systems; State and control constraints; Quadratic programming
\end{IEEEkeywords}

\IEEEpeerreviewmaketitle

\section{Introduction}\label{sec:1}

\IEEEPARstart{T}{he} reference governor (RG) is an add-on scheme to nominal closed-loop designs for handling pointwise-in-time constraints. It plays the role of a pre-filter with an adjustable bandwidth and modifies the evolution of input commands, when necessary, to enforce state and control constraints. The theory and applications of RGs and related schemes are covered in \cite{garone2017reference}, and references therein.

In this note, we describe a RG design for a pre-stabilized continuous-time
nonlinear system with the model
\begin{equation}\label{equ:sys_1}
\dot{x}(t) = f\big(x(t),v(t)\big) + w(t),
\end{equation}
where $x(t) \in \mathbb{R}^{n}$ is the state at time $t \in [0,\infty)$, $v(t) \in \mathbb{R}^{n_v}$ is the reference input at $t$, $w(t) \in \mathbb{R}^{n}$ is an unmeasured disturbance input at $t$, and the function $f: \mathbb{R}^n \times \mathbb{R}^{n_v} \rightarrow \mathbb{R}^n$ is nonlinear. The system must operate without violating the pointwise-in-time constraints given by
\begin{equation}\label{equ:constr_1}
x(t) \in X, \quad v(t) \in V, \quad \forall\, t \in [0,\infty).
\end{equation}

We make the following assumptions on $f$, $w(t)$, $X$, and $V$:

{\it (A1)} The solution of (\ref{equ:sys_1}) to any initial condition $x(0) = x_0 \in \mathbb{R}^{n}$, any piecewise continuous signal $v: [0,\infty) \to \mathbb{R}^{n_v}$, and any Lebesgue measurable signal $w: [0,\infty) \to \mathbb{R}^{n}$, exists and is unique on $[0,\infty)$.

We denote such a solution by $x(\cdot | x_0,v,w): [0,\infty) \to \mathbb{R}^{n}$. If $v \equiv \overline{v}$ for some constant $\overline{v} \in \mathbb{R}^{n_v}$, with a slight abuse of notation, we denote the solution corresponding to such a constant signal $v$ by $x(\cdot | x_0,\overline{v},w): [0,\infty) \to \mathbb{R}^{n}$.

{\it (A2)} The function $f(x,v)$ is continuously differentiable in $x$ and $v$. We use the notations $f_x = \frac{\partial f}{\partial x}$ and $f_v = \frac{\partial f}{\partial v}$.

{\it (A3)} The disturbance input signal $w: [0,\infty) \to \mathbb{R}^{n}$ is Lebesgue measurable and set-bounded as $w(t) \in W$ for all $t \in [0,\infty)$, where $W$ is a known compact set with $0 \in W$.

We denote the set of solutions of (\ref{equ:sys_1}) to a specific initial condition $x_0$, a specific piecewise continuous reference input signal $v$, and all disturbance input signals satisfying {\it (A3)} by $x(\cdot | x_0,v,W):= \{x(\cdot | x_0,v,w) \,|\, w \text{ satisfies {\it (A3)}} \}$.

{\it (A4)} Any constant reference input $\overline{v} \in \mathbb{R}^{n_v}$ associates with a unique globally asymptotically stable steady state $x_v(\overline{v})$, i.e., $f\big(x_v(\overline{v}),\overline{v}\big) = 0$ and $\lim_{t \to \infty} x(t | x_0,\overline{v},0) = x_v(\overline{v})$ for all $x_0 \in \mathbb{R}^{n}$.

{\it (A5)} The admissible sets $X$ and $V$ in \eqref{equ:constr_1} are both closed and convex. In this note, we assume them to be polyhedral: $X := \big\{ x \in \mathbb{R}^n \,|\, Mx \le m \big\}$ where $M \in \mathbb{R}^{n_m \times n}$, $m \in \mathbb{R}^{n_m}$, and $V := \big\{ v \in \mathbb{R}^{n_v} \,|\, M'v \le m' \big\}$ where $M' \in \mathbb{R}^{n_m' \times n_v}$, $m' \in \mathbb{R}^{n_m'}$.

The above assumptions can be relaxed. If {\it (A1)} does not hold globally, additional constraints may be added to $X$ to constrain $x(t)$ to the domain where existence and uniqueness of solutions hold. Similarly, if {\it (A4)} does not hold, for instance, steady states $x_v(\overline{v})$ are not unique or only regionally asymptotically stable, additional constraints may be used to constrain $x(t)$ to the domains of attraction of specific $x_v(\overline{v})$'s. The $X$ may need to be replaced with $X(\overline{v})$ to accomplish this, and this case can be handled in a similar way \cite{garone2016explicit}. Moreover, if $X$ or $V$ are not polyhedral, polyhedral subsets may be treated as the admissible sets.

The operation of a prediction-and-optimization based RG typically involves the following online optimization:

($\mathcal{P}1$) At each sample time $t$, solve
\begin{subequations}
\begin{align}
& v(t) = \argmin_{\overline{v} \in V}\,\, J(t) = \big\|\overline{v} - r(t)\big\|_S^2 \label{equ:RG_1} \\
& \quad\quad \text{s.t. }\,\, x \big(\tau \big| x(t),\overline{v},W\big) \subseteq X, \quad \forall\, \tau \in [0,\infty), \label{equ:constr_2}
\end{align}
\end{subequations}
where $r(t)$ is the reference command and $\|\cdot\|_S = \sqrt{(\cdot)^\top S (\cdot)}$ with $S = S^\top \succ 0$.

Hence, the RG adjusts the profile of $v(t)$ to satisfy \eqref{equ:constr_1} while maintaining $v(t)$ to be as close as possible to the reference command $r(t)$. A RG in this form is sometimes referred to as the command governor.

RG schemes for nonlinear systems have been pursued, e.g., in \cite{bemporad1998reference,angeli1999command,gilbert1999set,gilbert2002nonlinear,vahidi2007constraint}.
For general nonlinear systems, ($\mathcal{P}1$) is a nonlinear mathematical program with multiple constraints representing the constraint \eqref{equ:constr_2} imposed at all time instants over the prediction horizon and for all possible disturbance trajectories \cite{bemporad1998reference,angeli1999command}. This problem may not be easy to handle, especially when unmeasured disturbances are present. While alternatively level sets of input-to-state stability (ISS) Lyapunov functions may be employed to guard against constraint violation \cite{gilbert1999set}, ISS Lyapunov functions may not always be known and the use of level sets may, depending on the problem, lead to conservative/slow response. The explicit reference governor (ERG) \cite{garone2016explicit} is a non-optimization based scheme, where the constraints are enforced by a dynamic feedback law for $v(t)$. The ERG typically leads to slower response compared to a prediction-and-optimization based RG.

In this note, we consider another prediction-and-optimization based RG design for a continuous-time nonlinear system with an additive disturbance. This RG design exploits the prediction of the response of the nonlinear system based on the response of a linear model plus a prediction error. A state-and-input dependent bound on the prediction error is explicitly characterized using logarithmic norms \cite{dahlquist1958stability}, according to which the constraints are tightened. The online optimization ($\mathcal{P}1$) is then reduced to a convex quadratic program (QP) in ($\mathcal{P}3$), which can be easily and stably solved using standard QP solvers. While a similar strategy has been explored in our previous conference paper \cite{li2016reference}, there are substantial differences with the present note: 1) The design and analysis of most prediction-and-optimization based RG schemes, including that in \cite{li2016reference}, are based on discrete-time system models, while the RG design described in this note is for continuous-time systems. This avoids the errors in converting continuous-time models to discrete-time models and inter-sample constraint violations. 2) Differently from \cite{li2016reference} where parametrically convex set-valued mappings were used to over-bound the prediction error, we exploit logarithmic norms to characterize the error bound in this note. Such an approach has broader applicability, as a logarithmic norm can be induced from an arbitrary vector norm, thereby providing greater flexibility in the design.

To deal with a nonlinear system using an approximate linear model is attractive as predicting the state response of a linear model is computationally straightforward. However, the difference between the response of the nonlinear system and that predicted by the linear model should be compensated for, otherwise constraints may not be strictly enforced. A heuristic technique is to treat the difference between these responses as constant over the prediction horizon, but recursive feasibility may not be guaranteed \cite{vahidi2007constraint}. Another approach is to over-bound all such differences for the system operating in a certain range by a time-independent set and tighten the constraints according to this set bound. But such an approach may be conservative: the reference may respond slowly and not converge to steady-state constraint admissible commands \cite{li2016reference}. In contrast, advantages of the RG design using logarithmic norms described in this note include: 1) guaranteed constraint enforcement in the presence of unmeasured disturbances, including both sample-time and inter-sample constraint enforcement, 2) recursive feasibility, and 3) finite-time convergence of reference to command under appropriate assumptions.

\section{Preliminaries}\label{sec:2}

\subsection{Linearized model and error}

Consider a linearized model for \eqref{equ:sys_1} about an equilibrium state-input pair $\big(x_v(\overline{v}),\overline{v}\big)$,
\begin{equation}\label{equ:linear_1}
\delta \dot{x}(t) = f_x\big(x_v(\overline{v}),\overline{v}\big)\, \delta x(t) + f_v\big(x_v(\overline{v}),\overline{v}\big)\, \delta v(t) + w(t),
\end{equation}
where $\delta x(t)$ represents a predicted value of $x(t) - x_v(\overline{v})$ and $\delta v(t) := v(t) - \overline{v}$.

The steady state $x_v(\overline{v})$ plus the response of the linear model \eqref{equ:linear_1} gives an approximation of the response of the original nonlinear system \eqref{equ:sys_1}, but with an error, defined by
\begin{equation}\label{equ:error_1}
e(t) = x(t) - \big(x_v(\overline{v}) + \delta x(t)\big).
\end{equation}
% We derive a bound on this error exploiting logarithmic norms.

\subsection{Logarithmic norm}

An arbitrary vector norm $\|\cdot\|$ on $\mathbb{R}^n$ induces an operator norm $\|\cdot\|$ on linear operators $\mathbb{R}^{n} \to \mathbb{R}^{n}$ (identified by $\mathbb{R}^{n \times n}$), which in turn induces a real-valued functional $\mu: \mathbb{R}^{n \times n} \to \mathbb{R}$, called the ``logarithmic norm'' and defined by
\begin{equation}\label{equ:log_1}
\mu(F) = \lim_{h \rightarrow 0^+} \frac{\|I_n + h\, F \| - 1}{h}, \quad F \in \mathbb{R}^{n \times n}.
\end{equation}
Note that a ``logarithmic norm'' is not a norm on a vector space and can take negative values. For common vector norms, such as $\ell_p$-norms, $p=1,2,\infty$, their corresponding logarithmic norms admit explicit expressions \cite{afanasiev2013mathematical}. For instance, for a vector norm $\|\cdot\|_P = \sqrt{(\cdot)^\top P (\cdot)}$ with $P = P^\top \succ 0$, the corresponding logarithmic norm is given by
\small
\begin{equation}
\mu(F)= \lambda_{\max}
\bigg( \frac{(P^{1/2} F P^{-1/2})+(P^{1/2} F P^{-1/2})^\top }{2} \bigg),
\end{equation}
\normalsize
where $\lambda_{\max}(\cdot)$ represents the largest eigenvalue of a real symmetric matrix.

The basic result that links logarithmic norms to differential equations is as follows  \cite{dahlquist1958stability,soderlind2006logarithmic}:

{\it Proposition~1:} Consider $\dot{\theta}(t) = F\, \theta(t) + \gamma(t)$, where $F \in \mathbb{R}^{n \times n}$ and $\gamma: [0,\infty) \to \mathbb{R}^n$ is Lebesgue measurable. Then, (i) $D_t^+ \|\theta(t)\| \le \mu(F) \, \|\theta(t)\| + \|\gamma(t)\|$, where $D_t^+ \|\theta(t)\| := \lim_{h \rightarrow 0^+} \frac{\|\theta(t+h)\|-\|\theta(t)\|}{h}$. Furthermore, if $\mu(F)<0$ and $\|\gamma(t)\| \leq \gamma_{\max}$ for all $t \in [0,\infty)$, then, (ii) $\lim_{t \to \infty} dist \big(\theta(t),\Theta\big) = 0$ for all $\theta(0) \in \mathbb{R}^n$, and (iii) $\theta(0) \in \Theta \implies \theta(t) \in \Theta$ for all $t \in [0,\infty)$, where $\Theta:= \big\{ \theta \in \mathbb{R}^n  \,|\, \|\theta\|  \le -\frac{\gamma_{\max}}{\mu(F)} \big\}$ and $dist(\theta,\Theta):= \inf_{\theta_0 \in \Theta} \|\theta - \theta_0\|$.

{\em Proof:} See \cite{soderlind2006logarithmic}. $\blacksquare$

Proposition~1(ii) implies that $\mu(F)<0 \implies F$ is Hurwitz, i.e., eigenvalues of $F$ all have strictly negative real parts.

\section{Error bound based on logarithmic norms}\label{sec:3}

In this section, we exploit the logarithmic norm and Proposition~1 to derive a bound on the error \eqref{equ:error_1}. This bound is later used for the reference governor design.

The error \eqref{equ:error_1} is governed by the differential equation
\begin{align}
& \dot{e}(t) = \dot{x}(t) - \delta \dot{x}(t) \nonumber \\[3pt]
&= f \big(x(t),v(t) \big) - f_x\big(x_v(\overline{v}),\overline{v}\big)\, \delta x(t) - f_v\big(x_v(\overline{v}),\overline{v}\big)\, \delta v(t) \nonumber \\[4pt]
&= A(t)\, e(t) + B_x(t)\, \delta x(t) + B_v(t)\, \delta v(t), \label{equ:error_2}
\end{align}
where
\small
\begin{align*}
& A(t) := \int_{0}^{1} f_x \big(x_v(\overline{v}) + (e(t)+\delta x(t)) s,\overline{v}\big)\, ds, \\
& B_x(t) := \int_{0}^{1} \Big( f_x \big(x_v(\overline{v}) + (e(t)+\delta x(t)) s,\overline{v}\big) - f_x \big(x_v(\overline{v}),\overline{v}\big) \Big)\, ds, \\
& B_v(t) := \int_{0}^{1} \Big( f_v \big(x(t),\overline{v} + \delta v(t) s \big) - f_v \big(x_v(\overline{v}),\overline{v}\big) \Big)\, ds.
\end{align*}
\normalsize

{\it Proposition~2:} Given $\overline{v}$, suppose that (i) there exist $\mu_e = \mu_e(\overline{v}),\eta_x = \eta_x(\overline{v}),\eta_v = \eta_v(\overline{v}) \in \mathbb{R}$ such that
\begin{align}\label{equ:P21}
& \mu \big(f_x (\widehat{x},\overline{v})\big) \le \mu_e < 0, \nonumber \\[2pt]
& \big\| f_x (\widehat{x},\overline{v}) -f_x \big(x_v(\overline{v}),\overline{v}\big) \big\| \le \eta_x, \\[3pt]
& \big\| f_v (\widehat{x},\widehat{v}) -f_v \big(x_v(\overline{v}),\overline{v}\big) \big\| \le \eta_v, \nonumber
\end{align}
for all $\widehat{x} \in X$ and $\widehat{v} \in V$, and (ii) $\delta v(t) = \delta v$ and $\|w(t)\| \leq w_{\max}$ for all $t \in [0,\infty)$. Then, $\delta x(0) \in \Delta X(\overline{v},\delta v)$ and $e(0) \in E(\overline{v},\delta v)$ $\implies$ $e(t) \in E(\overline{v},\delta v)$ for all $t \in [0,\infty)$, where
\begin{align}
\Delta X(\overline{v},\delta v) \mkern-1mu &:= \mkern-1mu \big\{ \delta x \in \mathbb{R}^n \,|\, \|\delta x\| \le \Lambda_v(\overline{v}) \|\delta v\| + \Lambda_w(\overline{v}) w_{\max} \big\}, \label{equ:P221} \\
E(\overline{v},\delta v) \mkern-1mu &:= \mkern-1mu \big\{ e \in \mathbb{R}^n \,|\, \|e\| \le \Gamma_v(\overline{v}) \| \delta v\| + \Gamma_w(\overline{v}) w_{\max} \big\}, \label{equ:P222}
\end{align}
in which
\begin{align}
& \Lambda_v(\overline{v}) := - \frac{\|f_v(x_v(\overline{v}),\overline{v})\|}{\mu \big(f_x(x_v(\overline{v}),\overline{v})\big)}, \,\, \Lambda_w(\overline{v}) := - \frac{1}{\mu\big(f_x(x_v(\overline{v}),\overline{v})\big)}, \label{equ:P231} \\[-6pt]
& \Gamma_v(\overline{v}) := \frac{\eta_x \|f_v(x_v(\overline{v}),\overline{v})\| - \eta_v \mu\big(f_x(x_v(\overline{v}),\overline{v})\big)}{\mu_e \mu\big(f_x(x_v(\overline{v}),\overline{v})\big)}, \\
& \Gamma_w(\overline{v}) := \frac{\eta_x }{\mu_e \mu\big(f_x(x_v(\overline{v}),\overline{v})\big)}. \label{equ:P232}
\end{align}

{\em Proof:} By \eqref{equ:linear_1}, \eqref{equ:P221}, and Proposition~1(iii), $\delta x(0) \in \Delta X(\overline{v},\delta v)$ implies
\begin{equation}\label{equ:P25}
\|\delta x(t)\| \le  - \frac{\|f_v(x_v(\overline{v}),\overline{v})\|}{\mu \big(f_x(x_v(\overline{v}),\overline{v})\big)} \|\delta v\| - \frac{w_{\max}}{\mu \big(f_x(x_v(\overline{v}),\overline{v})\big)} ,
\end{equation}
where $\mu \big(f_x(x_v(\overline{v}),\overline{v})\big) \le \mu_e < 0$.

By \eqref{equ:error_2} and Proposition~1(i),
\begin{align}\label{equ:P26}
D_t^+ \|e(t)\| \le & \, \mu \big(A(t)\big) \|e(t)\| + \|B_x(t)\| \, \|\delta x(t)\| + \|B_v(t)\| \, \|\delta v \| \nonumber \\[3pt]
\le & \, \mu_e \, \|e(t)\| + \eta_x \, \|\delta x (t)\| + \eta_v \, \| \delta v \|.
\end{align}

Note that
\small
\begin{align*}
& \mu \big(A(t)\big) = \mu \Big(\int_{0}^{1} f_x \big(x_v(\overline{v}) + (e(t)+\delta x(t)) s,\overline{v}\big)\, ds \Big) \\
&\le \int_{0}^{1} \mu \Big(f_x \big(x_v(\overline{v}) + (e(t)+\delta x(t)) s,\overline{v}\big) \Big) ds \le \int_{0}^{1} \mu_e \, ds = \mu_e,
\end{align*}
\normalsize
by the fact that the logarithmic norm is sublinear \cite{soderlind2006logarithmic}. Similarly, $\|B_x(t)\| \le \eta_x$ and $\|B_v(t)\| \le \eta_v$.

Substituting \eqref{equ:P25} into \eqref{equ:P26}, we obtain
\begin{align*}%\label{equ:P27}
D_t^+ \|e(t)\| \le &\, \mu_e \, \|e(t)\| - \mu_e \Gamma_v (\overline{v})\, \|\delta v\| - \mu_e \Gamma_w(\overline{v})\, w_{\max}.
\end{align*}

Since $\mu_e<0$, by Proposition~1(iii), we obtain that $e(0) \in E(\overline{v},\delta v)$ implies $e(t) \in E(\overline{v},\delta v)$ for all $t \in [0,\infty)$.
$\blacksquare$

In the sequel, we make the following assumption:

{\it (A6)} For all $\overline{v} \in V$, the $\mu_e = \mu_e(\overline{v})$, $\eta_x = \eta_x(\overline{v})$, $\eta_v = \eta_v(\overline{v})$ defined in \eqref{equ:P21} exist. In particular, $\overline{\mu}_e := \sup_{\overline{v} \in V} \mu_e(\overline{v}) < 0$, $\overline{\eta}_x := \sup_{\overline{v} \in V} \eta_x(\overline{v}) < \infty$, and $\overline{\eta}_v:= \sup_{\overline{v} \in V} \eta_v(\overline{v}) < \infty$.

In the implementation, {\it (A6)} can be checked offline for a given closed-loop system. For instance, $\overline{\mu}_e$ can be estimated by solving the nonlinear program,
\begin{align}\label{equ:check_A6}
\max\, \mu \big(f_x (x,v)\big), \quad \text{s.t. } (x,v) \in X \times V,
\end{align}
using a global optimization algorithm (and similar for $\overline{\eta}_x$ and $\overline{\eta}_v$). Alternatively, {\it (A6)} can be enforced when the nominal stabilizing control for a given open-loop system is designed. For instance, suppose that the system is represented as $\dot{x}(t) = f\big(x(t),\pi(x(t),v(t),\rho)\big)$ where $\pi(\cdot,\cdot,\rho)$ is a control policy parameterized by $\rho$, then $\overline{\mu}_e < 0$ can be enforced by designing $\rho$ subject to the constraint,
\begin{align}\label{equ:enforce_A6}
\sup_{(x,v) \in X \times V} \mu \big(f_x (x,\pi(x,v,\rho))\big) \le \sigma,
\end{align}
for some $\sigma<0$. After {\it (A6)} is verified, a set of $\big\{(\mu_e,\eta_x,\eta_v)_i\big\}_{i=1}^{i_{\max}}$ satisfying \eqref{equ:P21} corresponding to a partition $\{V_i\}_{i=1}^{i_{\max}}$ of $V$ can be pre-computed offline and stored for online use. In particular, each $(\mu_e,\eta_x,\eta_v)_i$ satisfies \eqref{equ:P21} for all $\overline{v} \in V_i$, and is used when the current linearization point $\big(x_v(\overline{v}),\overline{v}\big)$ corresponds to $\overline{v} \in V_i$.

\section{Reference governor for nonlinear systems}\label{sec:4}

The reference governor (RG) updates the reference input at sample time instants $\{t_k\}_{k=0}^{\infty} \subset [0,\infty)$ (with $t_k \to \infty$ as $k \to \infty$), and maintains $v$ to be constant over each interval $[t_k,t_{k+1})$. As a result, the reference input signal $v$ is piecewise constant, and, by {\it (A1)} and {\it (A3)}, a solution of \eqref{equ:sys_1} exists and is unique. Let $v_k := v(t_k)$ be the RG output at the previous sample time instant, and $x_k := x_{v}(v_k)$ be the steady state corresponding to $v_k$. Denote the solution of \eqref{equ:linear_1} with $\big(x_v(\overline{v}),\overline{v}\big) = (x_k,v_k)$ and to the initial condition $\delta x(0) = x(t_{k+1}) - x_k$, the input signal $\delta v(t) \equiv \delta v$, and a specific disturbance input signal $w$, by $\delta x(\cdot|k,\delta v,w): [0,\infty) \to \mathbb{R}^n$. Let $e(t|k,\delta v,w) := x\big(t|x(t_{k+1}),v_k + \delta v,w\big) - \big(x_k + \delta x(t|k,\delta v,w)\big)$.

The constraint \eqref{equ:constr_2} can be written as
\begin{align}\label{equ:constr_4}
&\,\,\, x\big(t|x(t_{k+1}),v_k + \delta v,w\big) \nonumber \\
&= x_k + \delta x(t|k,\delta v,w) + e(t|k,\delta v,w) \in X \iff \nonumber \\
& \Big( \int_{0}^t \phi_{k}(t,\tau) f_v (x_k,v_k) \, d\tau \Big)\, \delta v + \int_{0}^t \phi_{k}(t,\tau) w(\tau)\, d\tau \nonumber \\
&+ e(t|k,\delta v,w) + x_k + \phi_{k}(t,0) \big(x(t_{k+1}) - x_k\big) \in X,
\end{align}
for all $t \in [0,\infty)$ and all $w$ satisfying {\it (A3)}, where
$\phi_{k}(t,\tau) = e^{f_x (x_k,v_k)(t-\tau)}$ is the state transition matrix associated with $f_x (x_k,v_k)$.

The variables $w(\tau)$ and $e(t|k,\delta v,w)$ are not measured but can be bounded: By {\it (A3)}, $w(\tau) \in W$. By Proposition~2, if
\begin{equation}\label{equ:Constra_Ini}
x(t_{k+1}) - x_k \in \Delta X(v_k,\delta v),
\end{equation}
where $\Delta X(\cdot,\delta v)$ is defined in \eqref{equ:P221} with $w_{\max} = \max_{w \in W} \|w\|$, then
\begin{equation}
e(t|k,\delta v,w) \in E(v_k,\delta v), \quad \forall\, t \in [0,\infty),
\end{equation}
where $E(\cdot,\delta v)$ is defined in \eqref{equ:P222}. Note that by {\it (A6)}, for each $k \in \mathbb{N}$, there exist $\mu_e(v_k)$, $\eta_x(v_k)$, and $\eta_v(v_k)$, used in the definitions of $\Delta X(v_k,\delta v)$ and $E(v_k,\delta v)$, satisfying \eqref{equ:P21}. Note also that $e(0|k,\delta v,w) = x(t_{k+1}) - \big(x_k + \delta x(0|k,\delta v,w)\big) = 0 \in E(v_k,\delta v)$.

Hence, the constraint \eqref{equ:constr_4} can be robustly enforced by enforcing the set inclusion
\begin{align}\label{equ:Constra_Time}
& \Big\{ \Big( \int_{0}^t \phi_{k}(t,\tau) f_v (x_k,v_k) \, d\tau \Big) \, \delta v \Big\} \oplus E(v_k,\delta v) \oplus F(v_k) \nonumber \\
& \subseteq X - \big(x_k + \phi_{k}(t,0) (x(t_{k+1}) - x_k) \big), \,\, \forall\, t \in [0,\infty),
\end{align}
where $\oplus$ denotes the Minkowski sum \cite{kolmanovsky1998theory}, and
\begin{align}\label{equ:W_1}
& F(v_k) := \big\{x \in \mathbb{R}^{n} \,|\, \|x\| \le \Lambda_w(v_k) \, w_{\max} \big\} \nonumber \\
& \supseteq \int_{0}^t \phi_{k}(t,\tau) W\, d\tau, \quad \forall\, t \in [0,\infty),
\end{align}
in which $\Lambda_w(\cdot)$ is defined in \eqref{equ:P231}, the integral is the Aumann integral \cite{aumann1965integrals}, and the set inclusion follows from Proposition~1.

We formulate the following online optimization for the RG:

($\mathcal{P}2$) At the sample time $t_{k+1}$, solve
\begin{subequations}\label{equ:RG_P2}
\begin{align}
& \delta v^* = \argmin_{\delta v \in \mathbb{R}^{n_v}}\,\, J(t_{k+1}) = \big\| v_k + \delta v - r(t_{k+1}) \big\|^2_S, \label{equ:RG_2} \\
&\quad\quad \text{s.t. }\,\,\,\, \delta v \in \Sigma_{k} \cup \{0\}, \label{equ:constr_3}
\end{align}
\end{subequations}
where
\begin{align}\label{equ:RG_P2_sigma}
\Sigma_{k} := \big\{ \delta v \in V - v_k \,|\,\,& \text{\eqref{equ:Constra_Ini} and \eqref{equ:Constra_Time} hold} \big\},
\end{align}
and output $v_{k+1} = v_k + \delta v^*$.

Note that if $f_x (x_k,v_k)$ is invertible, then $\int_{0}^t \phi_{k}(t,\tau)$ $f_v (x_k,v_k) \, d\tau =
\big(f_x (x_k,v_k)\big)^{-1} \big(\phi_{k}(t,0)-I_{n} \big)f_v (x_k,v_k)$.

\section{QP implementation and theoretical properties}\label{sec:5}

Difficulties in solving ($\mathcal{P}2$) are: 1) The constraint \eqref{equ:Constra_Time} is imposed on the continuous interval $t \in [0,\infty)$, i.e., at an uncountably infinite number of time instants. 2) The constraint \eqref{equ:Constra_Time} is in the form of set inclusion, which is in general not easy to handle. In this section, we transform ($\mathcal{P}2$) to a QP, and also establish theoretical results.

\subsection{QP implementation}\label{sec:51}

{\it Proposition~3:}
Suppose that for some $T \in [0,\infty)$ sufficiently large,
\begin{align}\label{equ:P3}
& \Big\{ \Big( \int_{0}^T \phi_{k}(T,\tau) f_v (x_k,v_k) \, d\tau \Big) \, \delta v \Big\} \oplus E(v_k,\delta v) \oplus F(v_k) \nonumber \\
& \subseteq \underline{X} - \big(x_k + \phi_{k}(T,0) (x(t_{k+1}) - x_k)\big),
\end{align}
where $\underline{X}$ is a closed set and $\underline{X} \subset \text{int}(X)$. Then, \eqref{equ:Constra_Time} holds for all $t \in [T,\infty)$.

{\em Proof:} Since $\underline{X}$ is closed and $\underline{X} \subset \text{int}(X)$, there exists $\epsilon>0$ such that $\underline{X} \oplus B(\epsilon) \subset X$, where $B(\epsilon):=\{x \in \mathbb{R}^n \,|\, \|x\| < \epsilon\}$. Let $\widehat{x}(t): = \phi_{k}(t,0) \big(x(t_{k+1}) - x_k\big) + \big( \int_{0}^t \phi_{k}(t,\tau) f_v(x_k,v_k) \, d\tau \big) \, \delta v$. Since $\mu \big(f_x (x_k,v_k)\big)< 0 \implies f_x (x_k,v_k)$ is Hurwitz, it follows that $\lim_{t \to \infty} \widehat{x}(t)  = \widehat{x}$ for some $\widehat{x} \in \mathbb{R}^n$. Then, given $\epsilon>0$, there exists $T \in [0,\infty)$, such that $\widehat{x}(t) - \widehat{x}(T) \in B(\epsilon)$ for all $t \in [T,\infty)$. The result of Proposition~3 follows. $\blacksquare$

{\it Proposition 4:} Suppose that given $t' \in [0,\infty)$ and $\Delta t > 0$,
\begin{align}\label{equ:P4}
& \Big\{ \Big( \int_{0}^{t'} \phi_{k}(t',\tau) f_v (x_k,v_k) \, d\tau \Big) \, \delta v \Big\} \oplus \widetilde{E}(v_k,\delta v,t',\Delta t) \nonumber \\
&  \oplus F(v_k) \subseteq X - \big(x_k + \phi_{k}(t',0) (x(t_{k+1}) - x_k)\big),
\end{align}
where
\begin{align}\label{equ:P41}
& \widetilde{E}(v_k,\delta v,t',\Delta t) := \big\{e \in \mathbb{R}^n \,|\, \|e\| \le \big( \Gamma_v(v_k) + \nonumber \\
&\quad \widetilde{\Gamma}_v(v_k,t',\Delta t) \big) \| \delta v\| + \Gamma_w(v_k) \, w_{\max} + \widetilde{\Gamma}_x(v_k,t',\Delta t) \big\}, \nonumber \\
& \widetilde{\Gamma}_v(v_k,t',\Delta t) := \| \phi_{k}(t',0) f_v (x_k,v_k) \|\, \xi(v_k,\Delta t), \nonumber \\
& \widetilde{\Gamma}_x(v_k,t',\Delta t) := \nonumber \\
&\quad\quad\quad \| f_x (x_k,v_k) \phi_{k}(t',0) \big(x(t_{k+1}) - x_k\big) \|\, \xi(v_k,\Delta t), \nonumber \\
& \xi(v_k,\Delta t): = \frac{\exp{\big(\Delta t\, \mu\big(f_x (x_k,v_k)\big)\big)}-1}{\mu\big(f_x (x_k,v_k)\big)}.
\end{align}
Then, \eqref{equ:Constra_Time} holds for all $t \in [t',t'+\Delta t]$.

{\em Proof:} For $t \in [t',t'+\Delta t]$, let $\widehat{x}(t): = \phi_{k}(t,0) \big(x(t_{k+1}) - x_k\big) + \big( \int_{0}^t \phi_{k}(t,\tau) f_v(x_k,v_k) \, d\tau \big) \, \delta v = \phi_{k}(t,t')\, \widehat{x}(t') + \big( \int_{t'}^t \phi_{k}(t,\tau) f_v(x_k,v_k) \, d\tau \big) \, \delta v$, which can be viewed as the state response of the system $\dot{\widehat{x}}(t) = f_x(x_k,v_k)\, \widehat{x}(t) + f_v(x_k,v_k) \, \delta v$ over $[t',t'+\Delta t]$ with the initial condition $\widehat{x}(t')$. Let $\Delta \widehat{x}(t) :=  \widehat{x}(t) - \widehat{x}(t')$, where $\widehat{x}(t')$ is treated as a constant. Then, $\Delta \dot{\widehat{x}}(t) = f_x(x_k,v_k) \, \big(\Delta \widehat{x}(t) + \widehat{x}(t')\big) + f_v(x_k,v_k) \, \delta v$ and $\Delta \widehat{x}(t') = 0$. Similar to \eqref{equ:P26}, for all $t \in [t',t'+\Delta t]$ we have
\begin{align}
& D_t^+ \|\Delta \widehat{x}(t)\| \le \mu\big(f_x(x_k,v_k)\big) \, \|\Delta \widehat{x}(t)\| \nonumber \\
&\quad\quad\quad\quad + \|f_x(x_k,v_k)\, \widehat{x}(t') + f_v(x_k,v_k) \, \delta v \| \implies \nonumber \\[2pt]
& \|\Delta \widehat{x}(t)\| \le \|f_x(x_k,v_k)\, \widehat{x}(t') + f_v(x_k,v_k) \, \delta v \| \nonumber \\[-2pt]
&\quad\quad\quad\quad \int_{t'}^{t} \exp{\big((t-\tau)\mu\big(f_x(x_k,v_k)\big) \big)}\, d\tau \nonumber \\
&\quad \le \|f_x(x_k,v_k)\, \widehat{x}(t') + f_v(x_k,v_k) \, \delta v \|\, \xi(v_k,\Delta t) \nonumber \\[-1pt]
&\quad \le \widetilde{\Gamma}_x(v_k,t',\Delta t) + \big\|f_x(x_k,v_k) \int_{0}^{t'} \phi_{k}(t',\tau) \nonumber \\[-2pt]
&\quad\quad f_v(x_k,v_k) \, d\tau  + f_v(x_k,v_k)\big\|\, \xi(v_k,\Delta t) \| \delta v \| \nonumber \\[4pt]
&\quad = \widetilde{\Gamma}_x(v_k,t',\Delta t) + \widetilde{\Gamma}_v(v_k,t',\Delta t)\, \| \delta v \|.
\end{align}
Then, $\big\{ \widehat{x}(t) \big\} \oplus E(v_k,\delta v) = \big\{ \widehat{x}(t') + \Delta \widehat{x}(t) \big\} \oplus E(v_k,\delta v) \subseteq \big\{ \widehat{x}(t') \big\} \oplus \widetilde{E}(v_k,\delta v,t',\Delta t)$. The result of Proposition~4 follows. $\blacksquare$

Note that $f_x (x_k,v_k)$ is Hurwitz $\implies \| \phi_{k}(t,0) f_v (x_k,v_k) \|$ and $\| f_x (x_k,v_k) \phi_{k}(t,0) \big(x(t_{k+1}) - x_k\big) \|$ are bounded on $t \in [0,\infty)$, and that $\lim_{\Delta t \to 0} \xi(v_k,\Delta t) = 0$. Thus, $\widetilde{\Gamma}_v(v_k,\cdot,\Delta t): [0,\infty) \to \mathbb{R}$ and $\widetilde{\Gamma}_x(v_k,\cdot,\Delta t): [0,\infty) \to \mathbb{R}$ converge uniformly to $0$ as $\Delta t \to 0$ .

Combining Propositions~3 and 4, we obtain:

{\it Corollary~1:} Suppose that
\begin{align}\label{equ:Constra_Time_sample}
& \Big\{ \Big( \int_{0}^{t} \phi_{k}(t,\tau) f_v (x_k,v_k) \, d\tau \Big) \delta v \Big\} \oplus \widetilde{E}(v_k,\delta v,t,\Delta t) \nonumber \\
& \oplus F(v_k) \subseteq \underline{X} - \big(x_k + \phi_{k}(t,0) (x(t_{k+1}) - x_k)\big)
\end{align}
holds at $t = 0, \Delta t, 2 \Delta t, \cdots, N \Delta t$ for some $N \in \mathbb{N}$ sufficiently large. Then, \eqref{equ:Constra_Time} holds for all $t \in [0,\infty)$.

The importance of Corollary~1 is that we only need to consider a finite number of constraints to enforce the uncountably infinite number of constraints \eqref{equ:Constra_Time} imposed on $[0,\infty)$.

In what follows, we introduce an approach to convert \eqref{equ:Constra_Time_sample} into linear inequalities and transform ($\mathcal{P}2$) to a QP.

By {\it (A5)}, \eqref{equ:Constra_Time_sample} can be written as
\begin{align}\label{equ:constr_8}
& \Big\{ M \Big( \int_{0}^{t} \phi_{k}(t,\tau) f_v (x_k,v_k) \, d\tau \Big) \delta v \Big\} \mkern-1mu \oplus \mkern-1mu M \mkern-1mu \widetilde{E}(v_k,\delta v,t,\Delta t) \nonumber \\
& \oplus \mkern-1mu M \mkern-1mu F(v_k) \le \underline{m} \mkern-1mu - \mkern-1mu Mx_k \mkern-1mu - \mkern-1mu M\phi_{k}(t,0) \big(x(t_{k+1}) - x_k\big),
\end{align}
where $\underline{m} := m - \delta$ with $\delta>0$ being arbitrarily small.

Suppose the support function for $\overline{B}:= \{x \in \mathbb{R}^n \,|\, \|x\| \le 1\}$, $h_{\overline{B}}(\cdot)$, is known. Then, given $M \in \mathbb{R}^{n_m \times n}$, we have
\begin{equation}
Mx = \begin{bmatrix} M_{1\cdot} x \\ \vdots \\M_{n_m\cdot} x \end{bmatrix} \le \begin{bmatrix} h_{\overline{B}}(M_{1\cdot}^\top) \\ \vdots \\ h_{\overline{B}}(M_{n_m\cdot}^\top) \end{bmatrix} =: H_{\overline{B}}(M),
\end{equation}
for all $x \in \overline{B}$, where $M_{i\cdot}$ denotes the $i$th row of $M$.

Then, \eqref{equ:constr_8} can be enforced by
\begin{align}\label{equ:constr_9}
& M \Big(\mkern-1mu \int_{0}^{t}\mkern-4mu \phi_{k}(t,\tau) f_v (x_k,v_k)\,\mkern-1mu d\tau \mkern-2mu\Big) \delta v \mkern-2mu + \mkern-2mu H_{1}(v_k,t,\Delta t,M) \|\delta v\| \nonumber \\
& \mkern-2mu + \mkern-2mu H_{2}(v_k,t,\Delta t,M) \le \underline{m} \mkern-2mu - \mkern-2mu Mx_k \mkern-2mu - \mkern-2mu M\phi_{k}(t,0) \big(x(t_{k+1}) \mkern-2mu - \mkern-2mu x_k\big),
\end{align}
where
\begin{align}\label{equ:constr_91}
H_{1}(v_k,t,\Delta t,M) :=&\, \big(\Gamma_v(v_k) + \widetilde{\Gamma}_v(v_k,t,\Delta t) \big) H_{\overline{B}}(M), \nonumber \\
H_{2}(v_k,t,\Delta t,M) :=&\, \Big( \big(\Gamma_w(v_k) + \Lambda_w(v_k) \big) \, w_{\max} \nonumber \\[-3pt]
&\, + \widetilde{\Gamma}_x(v_k,t,\Delta t)\Big) H_{\overline{B}}(M).
\end{align}

We constrain $\delta v$ to a polytope,
\begin{equation}\label{equ:constr_10}
\delta v \in \Delta V_{k}(\zeta) := \big\{ \delta v \in \mathbb{R}^{n_v} \,|\, U_{k} \, \delta v \le \zeta \, u_{k} \big\},\quad \zeta \ge 0,
\end{equation}
where $U_{k}$ and $u_{k}$ are specified, and $\zeta$ is an ancillary variable. Let $\overline{v}_k := \max_{\delta v \in \Delta V_{k}(1)} \|\delta v\|$, which can be easily computed for specified $U_{k}$ and $u_{k}$ by examining the vertices of $\Delta V_{k}(1)$. Then, we have
\begin{equation}
\max_{\delta v \in \Delta V_{k}(\zeta)} \|\delta v\| = \overline{v}_k \, \zeta,
\end{equation}
and \eqref{equ:constr_9} can be enforced by
\begin{align}\label{equ:constr_11}
& M \Big(\mkern-1mu \int_{0}^{t}\mkern-4mu \phi_{k}(t,\tau) f_v (x_k,v_k)\,\mkern-1mu d\tau \mkern-2mu\Big) \delta v \mkern-2mu + \mkern-2mu H_{1}(v_k,t,\Delta t,M)\, \overline{v}_k \, \zeta \nonumber \\
& \mkern-2mu + \mkern-2mu H_{2}(v_k,t,\Delta t,M) \le \underline{m} \mkern-2mu - \mkern-2mu Mx_k \mkern-2mu - \mkern-2mu M\phi_{k}(t,0) \big(x(t_{k+1}) \mkern-2mu - \mkern-2mu x_k\big),
\end{align}
which are linear inequalities in $(\delta v,\zeta)$.

We remark that, based on the cost function \eqref{equ:RG_2}, one possible choice for $(U_{k},u_{k})$ is of the form
\begin{equation}
U_{k} = \begin{bmatrix} I_{n_v} \\ -I_{n_v} \end{bmatrix} , \quad u_{k} = \left|\begin{bmatrix} S (v_k - r(t_{k+1})) \\ S (v_k - r(t_{k+1})) \end{bmatrix}\right|.
\end{equation}
This choice is motivated by considering the unconstrained steepest descent direction for $J(t_{k+1})$.

With the above manipulations, the online optimization for the RG is formulated as:

($\mathcal{P}3$) At the sample time $t_{k+1}$, solve
\begin{equation}\label{equ:RG_3}
\delta v^* = \argmin_{\delta v \in V - v_k}\,\, J(t_{k+1}) = \big\| v_k + \delta v - r(t_{k+1}) \big\|^2_S,
\end{equation}
subject to \eqref{equ:constr_10} and \eqref{equ:constr_11} for $t = 0, \Delta t, 2 \Delta t, \cdots, N \Delta t$ with $N \in \mathbb{N}$ sufficiently large. If a feasible solution $\delta v^*$ is found and satisfies \eqref{equ:Constra_Ini}, then output $v_{k+1} = v_k + \delta v^*$; otherwise, output $v_{k+1} = v_k$.

Note that the important feature of ($\mathcal{P}3$) is that the optimization involved in ($\mathcal{P}3$) is a QP, which can be easily solved. Note also that a structural difference between ($\mathcal{P}2$) and ($\mathcal{P}3$) is that the constraints \eqref{equ:Constra_Ini} and \eqref{equ:Constra_Time} are treated simultaneously in ($\mathcal{P}2$), and are treated sequentially in ($\mathcal{P}3$)\footnote{For the scalar reference case where $v(t) \in \mathbb{R}$, \eqref{equ:Constra_Ini} can be transformed to a linear inequality by restricting $\delta v \ge 0$ if $r(t_{k+1}) - v_k \ge 0$ and $\delta v \le 0$ if $r(t_{k+1}) - v_k \le 0$. Then, \eqref{equ:Constra_Ini} and \eqref{equ:Constra_Time} can be treated simultaneously and the optimization is still a QP.}.

\subsection{Theoretical properties}\label{sec:53}

{\it Proposition~5:} Suppose $x(t | x(0),v(0),W) \subseteq X$ for all $t \in [0,\infty)$. Then, (i) the optimization problems ($\mathcal{P}$2) and ($\mathcal{P}$3) are feasible at all sample time instants $\{t_k\}_{k=0}^{\infty}$. Let $v: [0,\infty) \to \mathbb{R}^{n_v}$ be the reference input signal generated by the RG through solving ($\mathcal{P}$2) or ($\mathcal{P}$3) at $\{t_k\}_{k=0}^{\infty}$. Then, (ii) $x(t | x(0),v,W) \subseteq X$ for all $t \in [0,\infty)$.

{\em Proof:} Property (i) holds trivially since $\delta v = 0$ is a feasible solution to ($\mathcal{P}$2) and ($\mathcal{P}$3) at all $\{t_k\}_{k=0}^{\infty}$. At each sample time instant $t_{k+1}$, if $x(t | x(t_{k}),v(t_k),W) \subseteq X$ for all $t \in [0,\infty)$, then $x(t | x(t_{k+1}),v(t_k) + \delta v,W) \subseteq X$ for all $t \in [0,\infty)$, where $\delta v$ is generated from ($\mathcal{P}$2) or ($\mathcal{P}$3). This is because: If $\delta v \neq 0$, the satisfaction of \eqref{equ:Constra_Ini} and \eqref{equ:Constra_Time} (or \eqref{equ:constr_10} and \eqref{equ:constr_11}) guarantees $x(t | x(t_{k+1}),v(t_k) + \delta v,W) \subseteq X$ for all $t \in [0,\infty)$; if $\delta v = 0$, $x(t | x(t_{k+1}),v(t_k),W) \subseteq x\big(t+(t_{k+1}-t_{k}) | x(t_{k}),v(t_k),W\big) \subseteq X$. The result of (ii) follows. $\blacksquare$

We now study the convergence property of $v(t)$ to $r(t)$. We make the following additional assumptions:

{\it (A7)} For any initial condition $x_0 \in \mathbb{R}^{n}$, any constant reference input $\overline{v} \in \mathbb{R}^{n_v}$, and any $\epsilon>0$, there exists $t' \in [0,\infty)$ such that
\begin{equation*}
x(t|x_0,\overline{v},W) \subseteq \big\{x_v(\overline{v})\big\} \oplus F(\overline{v}) \oplus \overline{B}(\epsilon), \quad \forall \, t \in [t', \infty),
\end{equation*}
where $\overline{B}(\epsilon)$ is the closure of $B(\epsilon)$.

The assumption {\it (A7)} is reasonable based on the global asymptotic stability of $x_v(\overline{v})$ (see {\it (A4)}) and the role of $F(\overline{v})$ defined in \eqref{equ:W_1} in bounding the effect of the disturbance input taking values in the bounded set $W$ (see {\it (A3)}).

{\it (A8)} The map $x_v: \overline{v} \mapsto x_v(\overline{v})$ is affine.

The assumption {\it (A8)} holds for many real-world systems, such as mechanical systems with generalized positions and velocities as states and with $v$ representing the generalized positions for the system to track. Examples satisfying {\it (A8)} include the system in Section~\ref{sec:6_2} and the ones given in references \cite{garone2016explicit,gilbert2002nonlinear,li2016reference}.

To guarantee the convergence of $v(t)$ to a proper constant command $r_s \in \mathbb{R}^{n_v}$ in finite time, we consider the following additional constraint sets:
\begin{align}
V_1 &:= \big\{v \in V \,\big|\, \{x_v(v)\} \oplus \overline{B}(\varepsilon) \subseteq X \big\}, \label{equ:Constra_ss_P2} \\
V_1' &:= \big\{v \in V \,\big|\, M x_v(v) + \varepsilon H_{\overline{B}}(M) \le m \big\}, \label{equ:Constra_ss_P3}
\end{align}
where $\varepsilon>0$ is specified and satisfies
\begin{equation}\label{equ:Constra_ss_epsilon}
\varepsilon> 2 \sup_{\overline{v} \in V} \big(\Gamma_w(\overline{v}) + \Lambda_w(\overline{v})\big) w_{\max}.
\end{equation}
Note that by {\it (A6)}, the right-hand side of \eqref{equ:Constra_ss_epsilon} is finite. Note also that $X$ and $V$ are convex (resp. polyhedral) (see {\it (A5)}) and $x_v$ is affine (see {\it (A8)}) $\implies$ $V_1$ is convex (resp. $V_1'$ is polyhedral).

In addition, we define
\begin{equation}\label{equ:Constra_cost}
V_2(\overline{v},r):=  \big\{v \in V \,\big|\, \|v - r\|^2_S \le \max\big(\| \overline{v} - r \|^2_S - \kappa,0\big) \big\},
\end{equation}
where $\kappa = \kappa(\varepsilon)>0$ is a sufficiently small constant.

{\it Proposition~6:} Suppose that (i) {\it (A7)} and {\it (A8)} hold, (ii) $v(0) \in V_1$ and $r(t) = r_s \in V_1$ for all $t \in [0,\infty)$, and add the requirement (iii) $v_{k} + \delta v \in V_1 \cap V_2\big(v_k,r(t_{k+1})\big)$ to the definition of the constraint set $\Sigma_{k}$ in \eqref{equ:RG_P2_sigma}. Then, there exists $t' \in [0,\infty)$ such that $v(t) = r_s$ for all $t \in [t',\infty)$, where $v: [0,\infty) \to \mathbb{R}^{n_v}$ is the reference input signal generated by the RG through solving ($\mathcal{P}$2) at $\{t_k\}_{k=0}^{\infty}$.

{\em Proof:} Construct the subsequence $\{t_{k_l}\}_{l=0}^{\infty} \subset \{t_k\}_{k=0}^{\infty}$ according to that $k_0 = 0$ and $t_{k_{l+1}}$ is the first element in $\{t_k\}_{k=0}^{\infty}$ after $t_{k_l}$ such that $v_{k_{l+1}} \neq v_{k_l}$. This way, $\{t_{k_l}\}_{l=0}^{\infty}$ is the sequence of time instants where the reference input signal $v$ has jumps. By (iii), the sequence of cost values $\{J(t_{k_l})\}_{l=0}^{\infty}$ is decreasing and bounded from below by $0$. Hence, $\{J(t_{k_l})\}_{l=0}^{\infty}$ converges to $\lim_{l \to \infty} J(t_{k_l}) =: \widehat{J}$. In particular, as $J(t_{0})$ is finite and $J(t_{k_{l+1}}) \le \max\big(J(t_{k_{l}}) - \kappa,0\big)$, there exists $l_0 \in \mathbb{N}$ such that $J(t_{k}) = J(t_{k_{l_0}}) = \widehat{J}$ and $v_{k} = v_{k_{l_0}} =: \widehat{v}$ for all $k \ge k_{l_0}$, i.e., $v$ converges to $\widehat{v}$ through a finite number of jumps. We now prove $\widehat{J} = 0$ and thus $\widehat{v} = r_s$ by contradiction.

Suppose that $\widehat{v} \neq r_s$. By {\it (A7)}, for any $\epsilon>0$, there exists $t' = t'(\epsilon) \in (t_{k_{l_0}},\infty) \cap \{t_k\}_{k=0}^{\infty}$ such that
\small
\begin{align}\label{equ:P2_01}
& x\big(t - t'|x(t'),\widehat{v},W\big) \subseteq x\big(t - t_{k_{l_0}}|x(t_{k_{l_0}}),\widehat{v},W\big) \nonumber \\
& \subseteq \big\{x_v(\widehat{v})\big\} \oplus F(\widehat{v}) \oplus \overline{B}\big(\Lambda_v(\widehat{v})\, \epsilon\big),
\end{align}
\normalsize
for all $t \in [t', \infty)$. It follows that
\small
\begin{equation}\label{equ:P2_02}
x\big(t - t'|x(t'),\widehat{v},W\big) - x_v(\widehat{v}) \subseteq \Delta X(\widehat{v},\epsilon u),
\end{equation}
\normalsize
for all $t \in [t', \infty)$, where $u$ is an arbitrary unit vector in $\mathbb{R}^{n_v}$, which, by the fact that $0 \in W$ and Proposition~2, implies
\small
\begin{align}\label{equ:P2_03}
& \pm\Big[x\big(t - t'|x(t'),\widehat{v},0\big) - \Big(x_v(\widehat{v}) + \widehat{\phi}(t-t',0) \big(x(t') - x_v(\widehat{v}) \big) + \nonumber \\[-4pt]
& \big(\int_{0}^{t-t'} \widehat{\phi}(t-t',\tau) f_v \big(x_v(\widehat{v}),\widehat{v}\big) \, d\tau \big)\, \epsilon u \Big)\Big] \in E(\widehat{v},\epsilon u),
\end{align}
\normalsize
for all $t \in [t', \infty)$, where $\widehat{\phi}(t,\tau):= e^{f_x (x_v(\widehat{v}),\widehat{v})(t-\tau)}$.

Combining \eqref{equ:P2_01}, \eqref{equ:P2_02}, and \eqref{equ:P2_03}, we obtain
\small
\begin{align}\label{equ:P2_04}
& \Big\{x_v(\widehat{v}) + \widehat{\phi}(t-t',0) \big(x(t') - x_v(\widehat{v}) \big) + \Big(\int_{0}^{t-t'} \widehat{\phi}(t-t',\tau) \nonumber \\[-4pt]
&  f_v \big(x_v(\widehat{v}),\widehat{v}\big) \, d\tau \Big)\, \epsilon u \Big\} \oplus E(\widehat{v},\epsilon u) \oplus F(\widehat{v}) \subseteq \big\{x\big(t - t'|x(t'),\widehat{v},0\big)\big\} \nonumber \\
& \oplus 2E(\widehat{v},\epsilon u) \oplus F(\widehat{v}) \subseteq \big\{x_v(\widehat{v})\big\} \oplus 2E(\widehat{v},\epsilon u) \oplus 2F(\widehat{v}) \nonumber \\
& \oplus \overline{B}\big(\Lambda_v(\widehat{v})\, \epsilon\big) = \big\{x_v(\widehat{v})\big\} \oplus \overline{B}\Big(2\big(\Gamma_w(\widehat{v})+\Lambda_w(\widehat{v})\big) w_{\max} \nonumber \\[-2pt]
& + \big(2 \Gamma_v(\widehat{v}) + \Lambda_v(\widehat{v})\big) \epsilon \Big) \subseteq X, \quad t \in [t', \infty),
\end{align}
\normalsize
for any $\epsilon \in (0,\frac{\varepsilon - 2(\Gamma_w(\widehat{v})+\Lambda_w(\widehat{v})) w_{\max}}{2 \Gamma_v(\widehat{v}) + \Lambda_v(\widehat{v})}]$, since $\widehat{v} \in V_1$. Let $\overline{\epsilon} := \frac{\varepsilon - 2(\Gamma_w(\widehat{v})+\Lambda_w(\widehat{v})) w_{\max}}{2 \Gamma_v(\widehat{v}) + \Lambda_v(\widehat{v})}$.

As $\widehat{v}, r_s \in V_1$ and $V_1$ is convex, any $\widehat{v} + \delta v = \lambda \widehat{v} + (1-\lambda) r_s \in V_1$, $\lambda \in [0,1]$. We now discuss two cases separately:

1) $0< \|\widehat{v} - r_s\| \le \overline{\epsilon}$. Let $\delta v = r_s - \widehat{v}$ with $\|\widehat{v} + \delta v - r_s\|^2_S = 0$, which satisfies \eqref{equ:Constra_Ini}, \eqref{equ:Constra_Time} for all $t \in [0,\infty)$ and $\widehat{v} + \delta v \in V_1 \cap V_2(\widehat{v},r_s)$ at the sample time instant $t' = t'(\|\widehat{v} - r_s\|)$.

2) $\|\widehat{v} - r_s\| > \overline{\epsilon}$. Let $\delta v = \frac{\overline{\epsilon}(r_s-\widehat{v})}{\|r_s-\widehat{v}\|}$ with $\|\delta v\| = \overline{\epsilon}$, which satisfies \eqref{equ:Constra_Ini}, \eqref{equ:Constra_Time} for all $t \in [0,\infty)$ and $\widehat{v} + \delta v \in V_1$ at the sample time instant $t' = t'(\overline{\epsilon})$. Furthermore,
\small
\begin{align*}
& \|\widehat{v} + \delta v - r_s\|^2_S = \Big(1-\frac{\overline{\epsilon}}{\|\widehat{v} - r_s\|}\Big)^2 \|\widehat{v} - r_s\|^2_S \\[-4pt]
& < \Big(1-\frac{\overline{\epsilon}}{\|\widehat{v} - r_s\|}\Big) \|\widehat{v} - r_s\|^2_S = \|\widehat{v} - r_s\|^2_S - \frac{\overline{\epsilon}\, \|\widehat{v} - r_s\|^2_S}{\|\widehat{v} - r_s\|} \\[2pt]
& < \|\widehat{v} - r_s\|^2_S - \overline{\epsilon}^2 c^2,
\end{align*}
\normalsize
where $c>0$ is a constant such that $\|\cdot\|_S \ge c\, \|\cdot\|$ by the equivalence of norms on $\mathbb{R}^{n_v}$. If $\kappa = \kappa(\overline{\epsilon})$ is selected such that
$\kappa \in (0, \overline{\epsilon}^2 c^2]$, then $\widehat{v} + \delta v \in V_2(\widehat{v},r_s)$.

In both cases, there exist $t' \in (t_{k_{l_0}},\infty) \cap \{t_k\}_{k=0}^{\infty}$ and $\delta v \in \mathbb{R}^{n_v}\setminus\{0\}$ such that $\widehat{v} + \delta v$ is a feasible solution to ($\mathcal{P}$2) at $t'$ and has lower cost value than $\widehat{v}$. Then, there exists an optimal feasible solution to ($\mathcal{P}$2) at $t'$, $v^* \neq \widehat{v}$, having lower cost value than $\widehat{v}$, which contradicts the assumption that $v_{k} = \widehat{v} \neq r_s$ for all $k \ge k_{l_0}$. Therefore, $\widehat{v} = r_s$.

Finally, the above also shows that for every $t \in [0,\infty)$ where $v(t) \neq r_s$, $v$ has a jump at some $t' \in [t,\infty)$. Since $v: [0,\infty) \to \mathbb{R}^{n_v}$ converges to $\widehat{v} = r_s$ through a finite number of jumps, the result of Proposition~6 follows. $\blacksquare$

{\it Corollary~2:} Suppose that {\it (A7)}, {\it (A8)} hold, $v(0) \in V_1'$, and $r(t) = r_s \in V_1'$ for all $t \in [0,\infty)$. Solve the QP \eqref{equ:RG_3} with sufficiently small $\Delta t>0$ and additional linear inequality constraints corresponding to $v_{k} + \delta v \in V_1'$. If a feasible solution $\delta v^*$ is found and satisfies \eqref{equ:Constra_Ini} and $v_k + \delta v^* \in V_2\big(v_k,r(t_{k+1})\big)$, then output $v_{k+1} = v_k + \delta v^*$; otherwise, output $v_{k+1} = v_k$. Then, there exists $t' \in [0,\infty)$ such that $v(t) = r_s$ for all $t \in [t',\infty)$, where $v: [0,\infty) \to \mathbb{R}^{n_v}$ is the reference input signal generated by the RG through solving ($\mathcal{P}$3) at $\{t_k\}_{k=0}^{\infty}$.

{\em Sketch of Proof:} Due to the structural difference between ($\mathcal{P}$2) and ($\mathcal{P}$3), the proof of Corollary~2 is slightly different from that of Proposition~6.

The existence of a feasible solution $\delta v$ satisfying \eqref{equ:Constra_Ini}, \eqref{equ:constr_10}, \eqref{equ:constr_11}, and $\widehat{v} + \delta v \in V_1' \cap V_2(\widehat{v},r_s)$ at $t'$, where $t'$ is defined similarly to that in the proof of Proposition~6, guarantees the existence of an optimal feasible solution $\delta v^*$ to the QP \eqref{equ:RG_3} (augmented with $v_{k} + \delta v \in V_1'$), which must satisfy $\widehat{v} + \delta v^* \in V_2(\widehat{v},r_s)$.

If $\|\delta v^*\| \ge \overline{\epsilon}$, $\delta v^*$ satisfies \eqref{equ:Constra_Ini} and the RG outputs $v(t') = \widehat{v} + \delta v^*$. If $\|\delta v^*\| < \overline{\epsilon}$, the RG may reject $\delta v^*$ when checking the condition \eqref{equ:Constra_Ini} and output $v(t') = \widehat{v}$. In this case, based on a similar expression as \eqref{equ:P2_04}, the constraints \eqref{equ:constr_10}, \eqref{equ:constr_11}, and $\widehat{v} + \delta v^* \in V_1'$ are inactive at $t'$ and remain inactive for all $t \in [t',\infty)$. Thus, $\delta v^*$ remains the minimizer of \eqref{equ:RG_3} for all $t_k \ge t'$. By {\it (A7)}, there exists $t'' \in [t',\infty) \cap \{t_k\}_{k=0}^{\infty}$ such that
\small
\begin{align}
& x(t'') - x_v(\widehat{v}) \in x\big(t'' - t'|x(t'),\widehat{v},W\big) - x_v(\widehat{v}) \nonumber \\[1pt]
& \subseteq F(\widehat{v}) \oplus \overline{B}\big(\Lambda_v(\widehat{v}) \|\delta v^*\| \big) = \Delta X(\widehat{v},\delta v^*).
\end{align}
\normalsize
Consequently, the RG outputs $v(t'') = \widehat{v} + \delta v^*$ at $t''$. The result of Corollary~2 then follows from similar augments as those in the proof of Proposition~6. $\blacksquare$

Note that in the disturbance-free case, $W = \{0\}$, the $\varepsilon>0$ in the definitions of $V_1$ and $V_1'$ can be arbitrarily small. Thus, $V_1$ and $V_1'$ represent the set of all reference commands whose corresponding steady states locate in the interior of the state admissible set $X$. In turn, Proposition~6 and Corollary~2 guarantee the finite-time convergence of the reference to any strictly steady-state constraint admissible commands.

\section{Examples}\label{sec:6}

\subsection{Second-order nonlinear system with a scalar reference}\label{sec:6_1}

We consider the following nonlinear system,
\begin{align}\label{equ:Ex1}
\dot{x}_1 &= -0.5 \sin x_1 + x_2 + 0.5 v + w, \nonumber \\
\dot{x}_2 &= -\sin x_1 - 1.5 x_2 + v,
\end{align}
where $(\overline{x}_1,0)$ is the steady state corresponding to the constant reference input $\overline{v} = \sin \overline{x}_1$, and $w$ is an unmeasured disturbance input taking values in $[-w_{\max},w_{\max}]$ based on a truncated Gaussian distribution. The imposed state constraint set is $X = \left[-\frac{\pi}{4},\frac{\pi}{4}\right] \times [-0.2,0.2]$.

In the design of our RG, we use the Euclidean norm $\|\cdot\|$ and its corresponding logarithmic norm $\mu(\cdot)$. In this example, $f_x(x,v) = \begin{bmatrix} - 0.5 \cos x_1 & 1 \\ - \cos x_1 & -1.5 \end{bmatrix}$ only depends on scalar $x_1$ and $f_v(x,v) = [0.5, 1]^{\top}$ is constant. Thus, $\mu_e$ and $\eta_x$ are easy to estimate and $\eta_v \equiv 0$. In particular, we partition the interval $\left[-\frac{\pi}{4},\frac{\pi}{4}\right]$ and estimate $\mu_e$ and $\eta_x$ (see \eqref{equ:P21}) by examining $x_1$'s on the grid points at each sample time.

Apart from implementing our RG for nonlinear systems (RG-NL), we also implement a nonlinear-program based RG (RG-NP), i.e., predicting the state responses by directly simulating the nonlinear system \eqref{equ:Ex1} and imposing the constraints at all sample time instants over a sufficiently long prediction horizon. Furthermore, we implement a RG for linear systems (RG-L) to treat the nonlinear system \eqref{equ:Ex1}, i.e., based on a linear model in the form of \eqref{equ:linear_1} with compensation for the disturbance $w$ but without that for the linear model prediction errors. The sampling frequency is 20 [Hz] for all the designs.

We test the case where the reference command is $r_s = \sin (\frac{\pi}{4})$ so that the desired steady state $x_v(r_s) = (\frac{\pi}{4},0)$ is at the boundary of the set $X$. We test two cases for the unmeasured disturbance, $w_{\max} = 1 \times 10^{-2}$ and $w_{\max} = 0$.

%%%%%%%%%%%%%%%%%%%%%%%%%%%%%%%%%%%%%%%%%%%%%%%%%%%%%%%%%%%%%
\begin{figure}[h!]
\begin{center}
\begin{picture}(360.0, 348.0)
\put(  0,  224){\epsfig{file=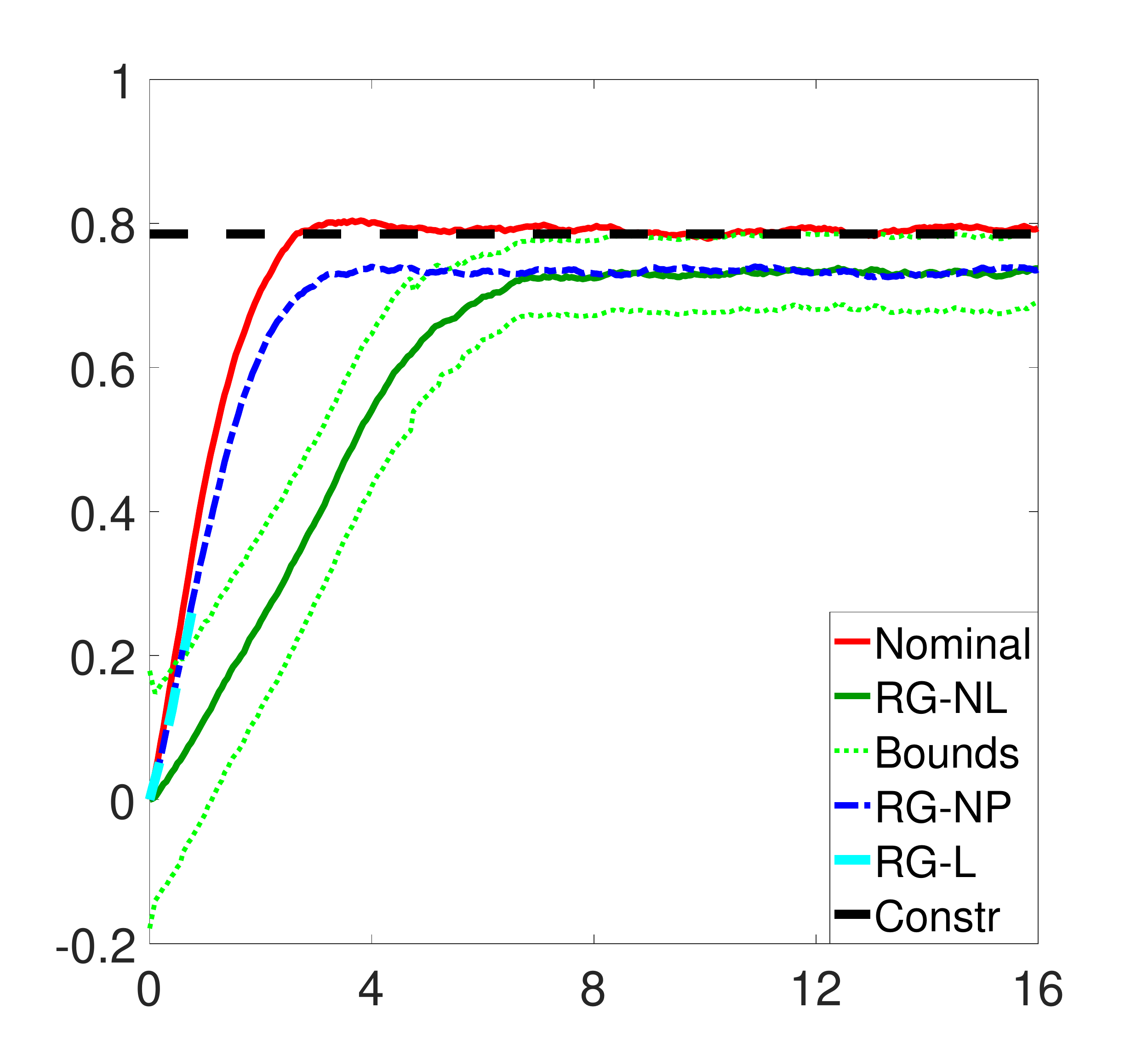,width=1.8in}}  %%%
\put(  0,  109){\epsfig{file=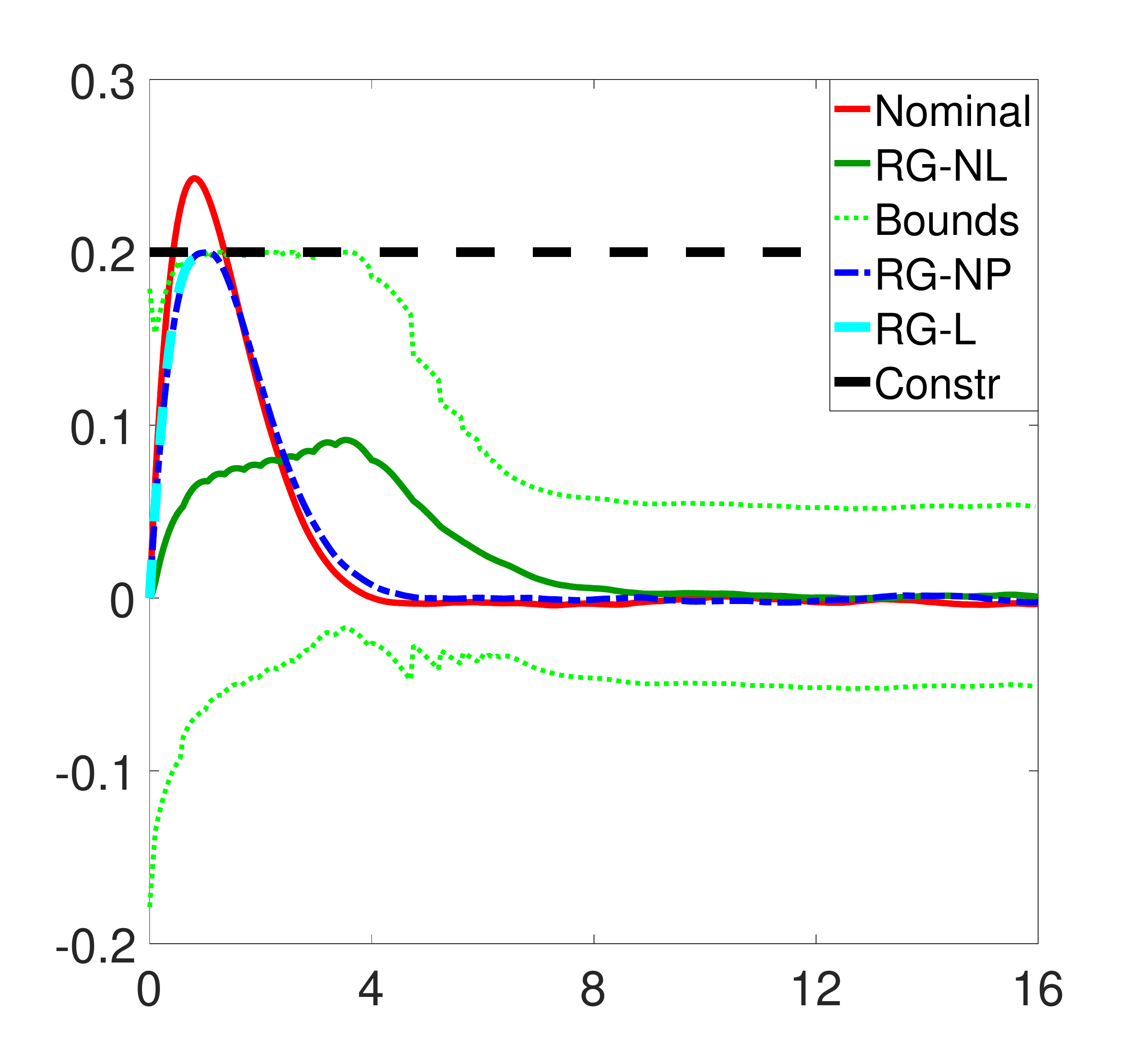,width=1.8in}}  %%%
\put(  0,  -6){\epsfig{file=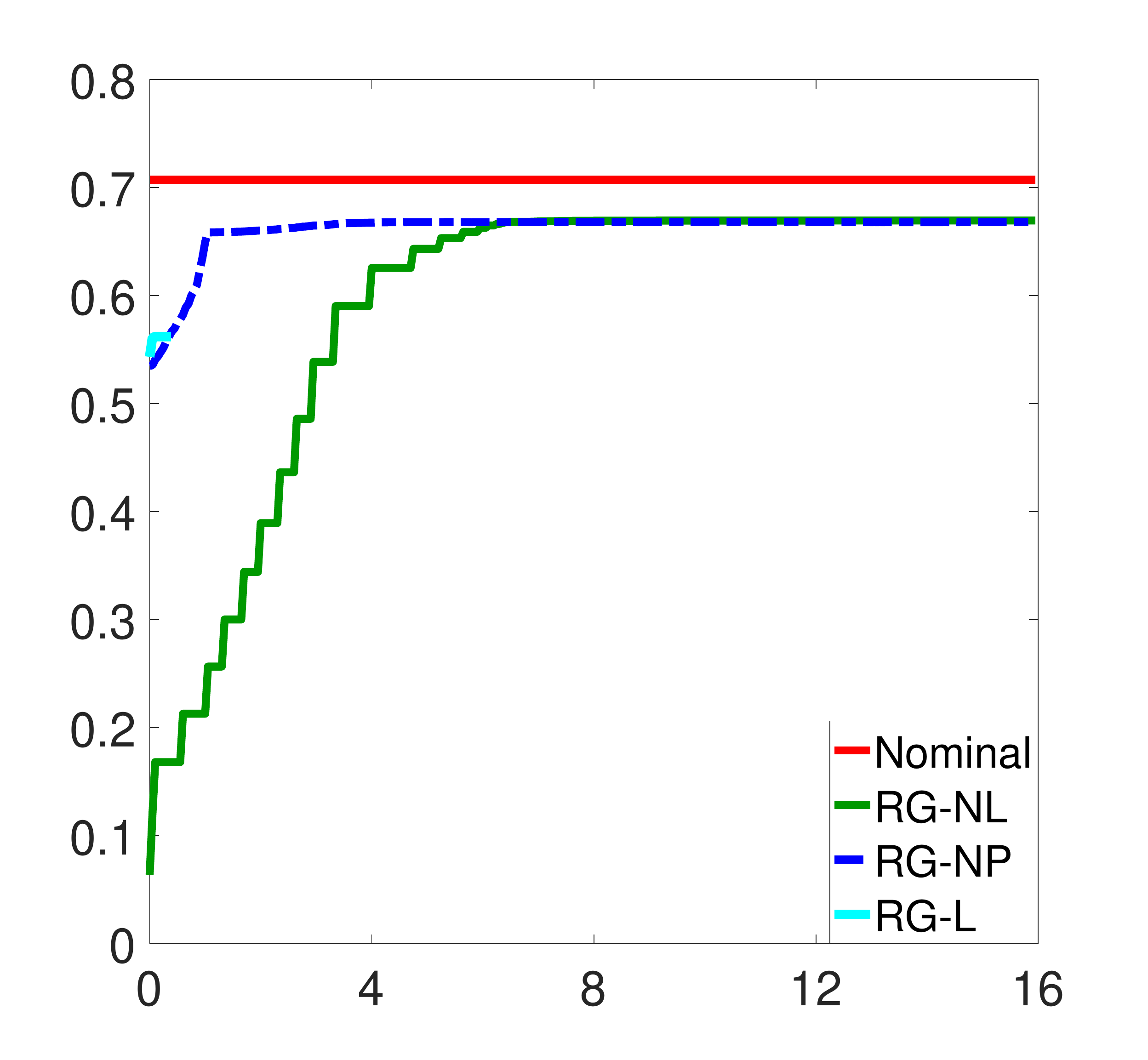,width=1.8in}}  %%%
\put(  105,  -4){$t$}  %%%
\put(  105,  111){$t$}  %%%
\put(  105,  226){$t$}  %%%
\put(  0,  94){$v$}  %%%
\put(  0,  209){$x_2$}  %%%
\put(  0,  326){$x_1$}  %%%

\put(  130,  224){\epsfig{file=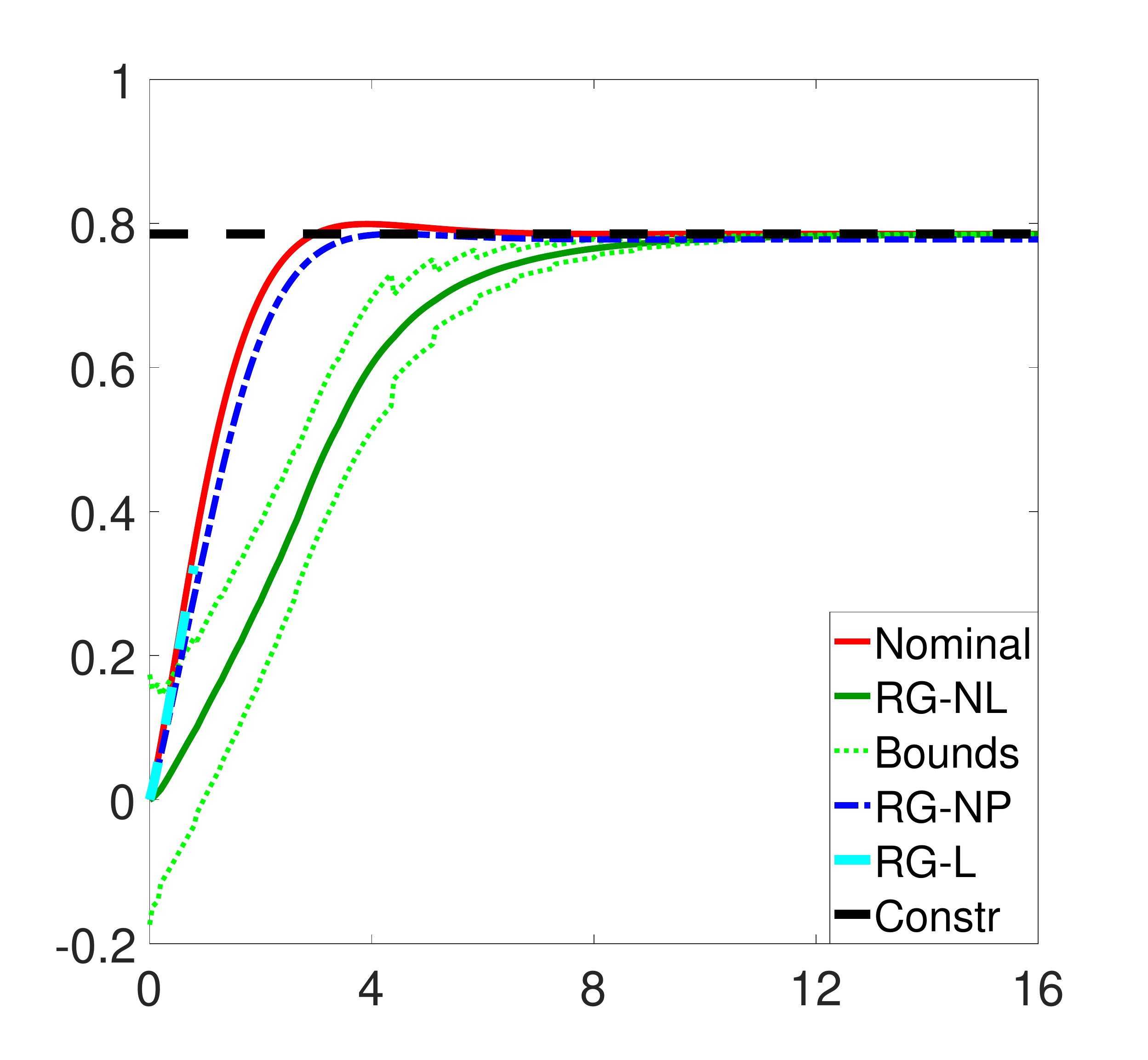,width=1.8in}}  %%%
\put(  130,  109){\epsfig{file=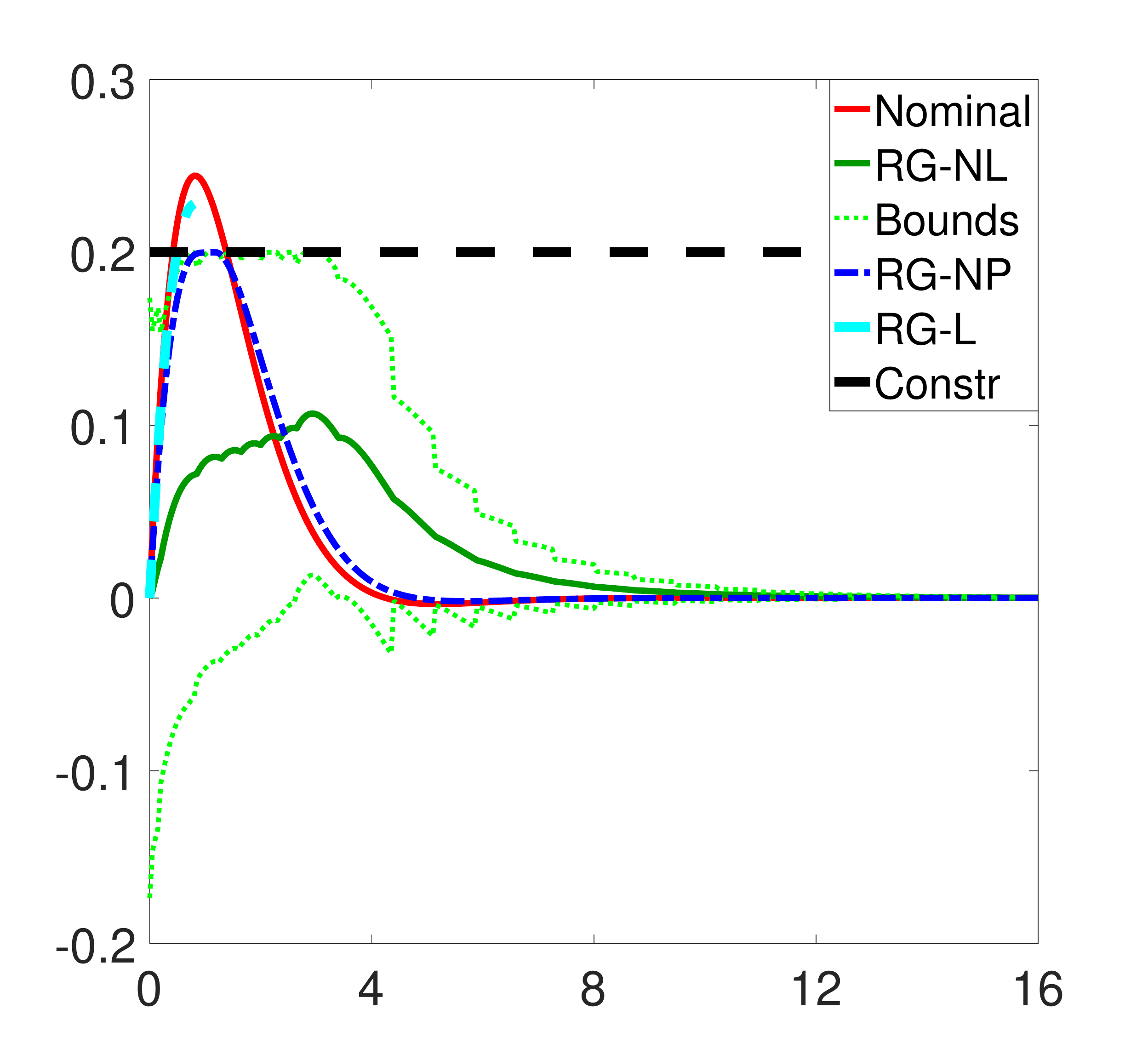,width=1.8in}}  %%%
\put(  130,  -6){\epsfig{file=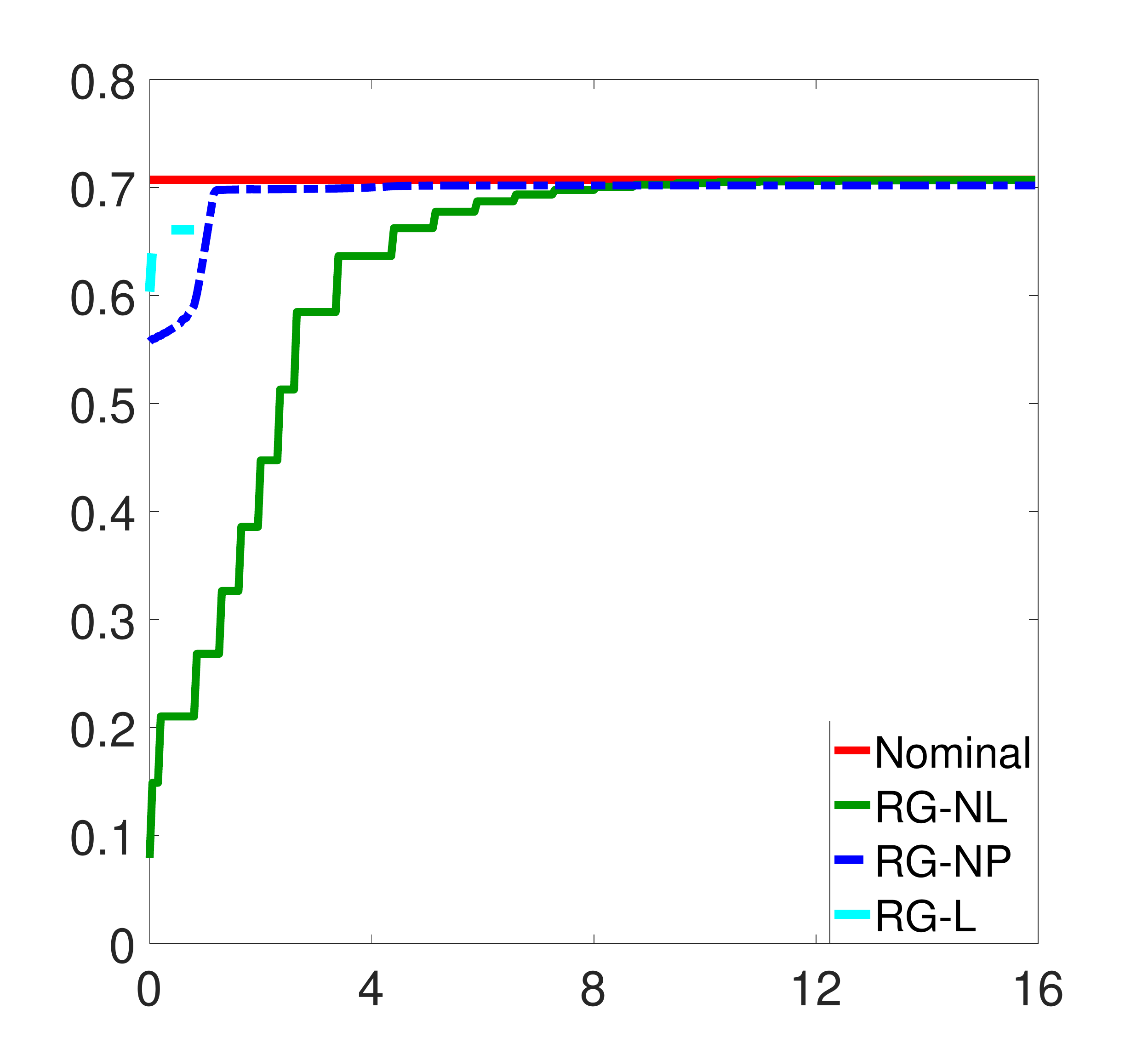,width=1.8in}}  %%%
\put(  235,  -4){$t$}  %%%
\put(  235,  111){$t$}  %%%
\put(  235,  226){$t$}  %%%
\put(  130,  94){$v$}  %%%
\put(  130,  209){$x_2$}  %%%
\put(  130,  326){$x_1$}  %%%

\put(  104,  339){(a)}  %%%
\put(  234,  339){(b)}  %%%

\end{picture}
\end{center}
      \caption{Results for (a) $w_{\max} = 1 \times 10^{-2}$, and (b) $w_{\max} = 0$.}
      \vspace{-0.15in}
      \label{fig:Ex_1}
\end{figure}
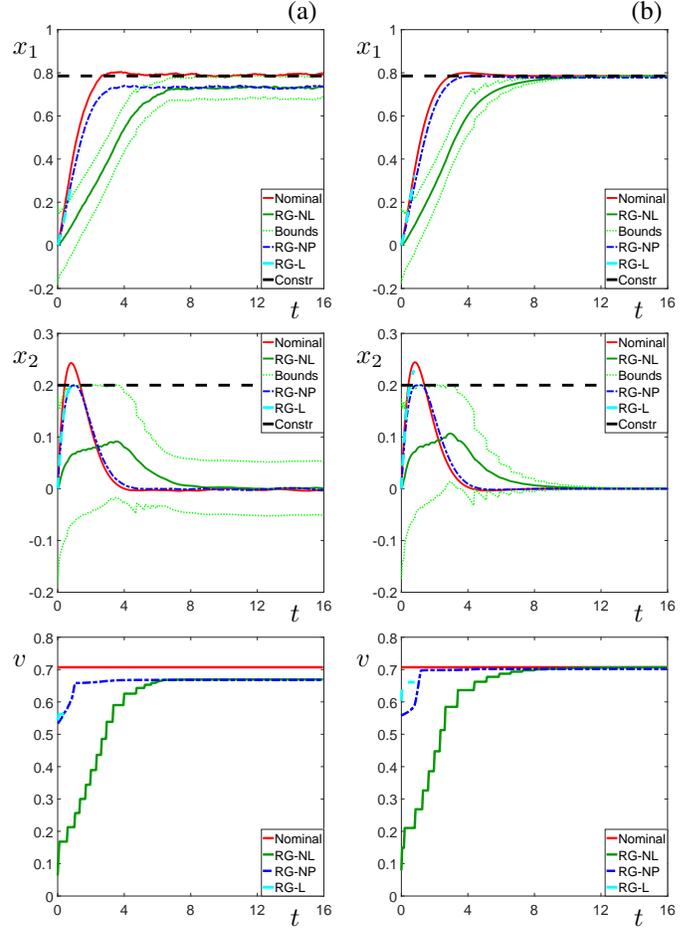
%%%%%%%%%%%%%%%%%%%%%%%%%%%%%%%%%%%%%%%%%%%%%%%%%%%%%%%%%%%%%

The simulation results are shown in Fig.~\ref{fig:Ex_1}, where the red solid curves represent the state/reference responses of the nominal system without a RG, the dark-green solid curves represent the responses using RG-NL, the light-green dotted curves represent the bounding sets of the predicted state responses, the dark-blue dash-dotted curves represent the responses using RG-NP, the light-blue dashed curves represent the responses using RG-L, and the horizontal black dashed lines represent the state constraints.

Without a RG, the state responses violate the constraints in both cases. A RG based only on a linear model fails to guard against constraint violations. In Fig.~\ref{fig:Ex_1}(a), for some future trajectory of $w$ taking values in $[-w_{\max},w_{\max}]$, the constraint on $x_2$ will be violated and there is no solution for the RG to avoid it so the solver fails. In Fig.~\ref{fig:Ex_1}(b), the constraint on $x_2$ is violated. Then the simulations get stopped. Using RG-NL or RG-NP, the constraints are strictly enforced. In Fig.~\ref{fig:Ex_1}(a), when unmeasured disturbances are present, RG-NL and RG-NP drive $v(t)$ to almost the same value such that a safety margin is kept between the converged state and the boundary of $X$. In Fig.~\ref{fig:Ex_1}(b), when no disturbance is present, both RG-NL and RG-NP drive $v(t)$ to $r_s$. Note that although the reference response is slower using RG-NL than using RG-NP, RG-NL requires much lower computational effort than RG-NP. The average CPU time at each sample time, including the time to estimate $\mu_e$ and $\eta_x$ and that to solve for a new reference value using Matlab {\it quadprog}, is 5.5~[ms] for RG-NL, as opposed to 172.7~[ms] for RG-NP using Matlab {\it fmincon}, both with an Intel Core i7-4790 3.60 GHz processor, 16.0 GB RAM, and solvers' default settings. Note that although both computation times could be improved through code optimization and implementation in C, RG-NL is more than 30 times faster than RG-NP in our implementation.

\subsection{Spacecraft attitude control}\label{sec:6_2}

We consider the attitude control of a spacecraft, with open-loop dynamic equations
\begin{align}
\begin{bmatrix} \dot{\phi} \\ \dot{\theta} \\ \dot{\psi} \\ \dot{\omega}_1 \\ \dot{\omega}_2 \\ \dot{\omega}_3 \end{bmatrix} &= \begin{bmatrix} \frac{1}{\cos{\theta}} \begin{bmatrix} \cos{\theta} & \sin{\phi} \sin{\theta} & \cos{\phi}\sin{\theta} \\ 0 & \cos{\phi} \cos{\theta} & - \sin{\phi} \cos{\theta} \\ 0 & \sin{\phi} & \cos{\phi} \end{bmatrix} \begin{bmatrix} \omega_1 \\ \omega_2 \\ \omega_3 \end{bmatrix} \\
 \frac{1}{J_1} (J_2 - J_3) \omega_2 \omega_3 + \frac{M_1}{J_1} + w_1 \\
 \frac{1}{J_2} (J_3 - J_1) \omega_3 \omega_1 + \frac{M_2}{J_2} + w_2 \\
 \frac{1}{J_3} (J_1 - J_2) \omega_1 \omega_2 + \frac{M_3}{J_3} + w_3 \end{bmatrix},
\end{align}
where $M_{1,2,3}$ are torques provided by thrust or power, and $w_{1,2,3}$ are external disturbances.

The nominal controller is a linear quadratic regulator (LQR) designed based on the linearized model at the origin $x = [\phi,\theta,\psi,\omega_1,\omega_2,\omega_3]^{\top} =[0,0,0,0,0,0]^{\top}$, i.e., \\
\begin{equation}
\dot{x} = (A-BK) x + BK \big[v_1, v_2, v_3, 0,\, 0,\, 0 \big]^{\top},
\end{equation}
where $v_{1,2,3}$ are reference inputs representing the desired steady-state orientation angles, i.e., $[\phi_v(v_1),\theta_v(v_2),\psi_v(v_3)] = [v_1,v_2,v_3] = [\phi_{\text{des}},\theta_{\text{des}},\psi_{\text{des}}]$;
\begin{equation}
A = \begin{bmatrix} 0_{3 \times 3} & I_{3 \times 3} \\ 0_{3 \times 3} & 0_{3 \times 3} \end{bmatrix}, \quad B = \begin{bmatrix} 0_{3 \times 3} \\ \text{\small$\begin{bmatrix} 1/J_1 & 0 & 0 \\ 0 & 1/J_2 & 0 \\ 0 & 0 & 1/J_3 \end{bmatrix}\normalsize$} \end{bmatrix};
\end{equation}
and $K$ is obtained by solving the continuous-time algebraic Riccati equation,
\begin{equation}
A^\top P + P A - P B R^{-1} B^\top P + Q = 0, \quad K = R^{-1} B^{\top} P,
\end{equation}
to stabilize the system.

The spacecraft and controller parameter values are: $J_1 =$ 120 [kg$\cdot$m$^2$], $J_2 =$ 100 [kg$\cdot$m$^2$], $J_3 =$ 80 [kg$\cdot$m$^2$], $Q =$ diag($1$), and $R =$ diag($1\times 10^{-3}$). We assume that the disturbances $[w_1,w_2,w_3]^{\top} \in W$, where $\max_{w \in W} \|w\| = 2 \times 10^{-3}$.

The objective is to steer the spacecraft from the initial steady state corresponding to $v(0)$ to a desired steady state corresponding to $r_s$, where
\begin{equation}
v(0) = \big[-\frac{\pi}{18}, -\frac{\pi}{20}, -\frac{\pi}{24} \big]^\top, \quad r_s = \big[\frac{\pi}{20}, \frac{\pi}{20}, \frac{\pi}{20} \big]^\top,
\end{equation}
while satisfying the following constraints for all $t \in [0,\infty)$,
\begin{equation}\label{equ:Ex2_bd}
|\phi(t)| \le 0.2, \quad
|\theta(t)| \le 0.2, \quad
|\psi(t)| \le 0.2.
\end{equation}

This problem is challenging as: 1) It is in general difficult to explicitly characterize the set of reachable states of a nonlinear system under disturbances, especially when the dimensions of state and disturbance are both higher than $1$. As a result, a nonlinear-program based RG may not be easy to implement. 2) A spacecraft typically has limited computing power, requiring the control scheme to have low computational complexity. Our RG scheme is a suitable choice in this case.

Motivated by the fact that $V(\cdot) = (\cdot)^\top P (\cdot)$ is a Lyapunov function for the LQR closed-loop system, we use the logarithmic norm corresponding to the vector norm $\|\cdot\|_P = \sqrt{(\cdot)^\top P (\cdot)}$ for our RG design. It is verified via solving the nonlinear program \eqref{equ:check_A6} offline that using this logarithmic norm, {\it (A6)} holds for the operation range of the spacecraft. In particular, $f_x(x,v)$ is independent of $v$ and satisfies $\mu\big(f_x(x,v)\big) \le -0.261$ for all $x = [\phi,\theta,\psi,\omega_1,\omega_2,\omega_3]^{\top}$ satisfying the constraints \eqref{equ:Ex2_bd} and $|\omega_i| \le 0.05$, $i = 1,2,3$, and $f_v(x,v)$ is constant. We set $\mu_e \equiv -0.261$, $\eta_v \equiv 0$, and estimate $\eta_x$ by examining vertices of the box defined by \eqref{equ:Ex2_bd} and $|\omega_{1,2,3}| \le 0.05$ at each sample time.

\begin{figure}[h!]
\begin{center}
\begin{picture}(360, 118.0)
\put(  0,  -6){\epsfig{file=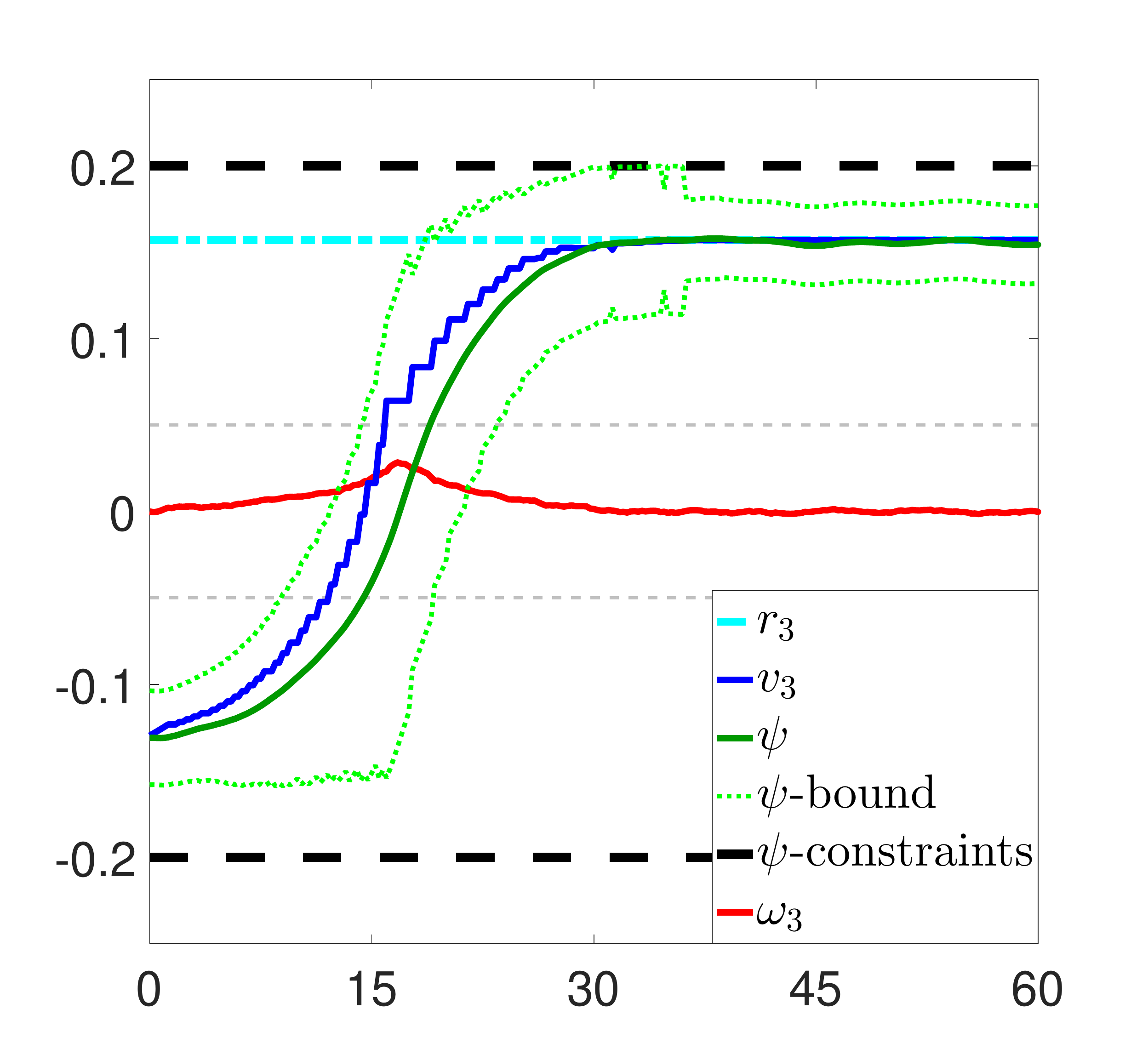,width=1.8in}}
\put(  130,  -6){\epsfig{file=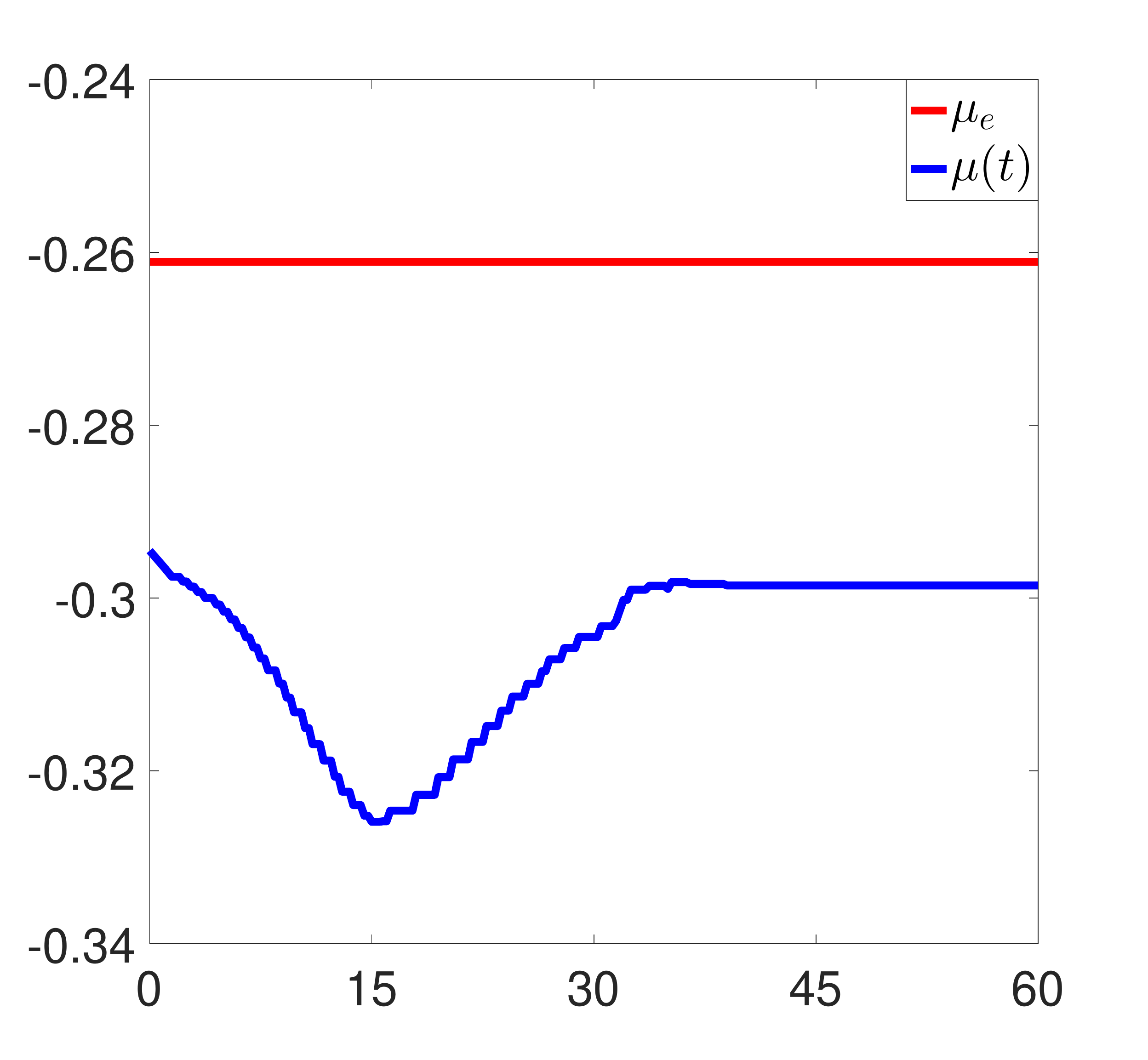,width=1.8in}}
\put(  96,  -6.5){$t$ [s]}
\put(  226, -6.5){$t$ [s]}
\put(  0,  109){$\psi$ [rad] or $\omega_3$ [rad/s]}

\put(  104,  109){(a)}  %%%
\put(  234,  109){(b)}  %%%

\end{picture}
\end{center}
      \caption{Results for spacecraft attitude control.}
      \vspace{-0.15in}
      \label{fig:Ex_2}
\end{figure}
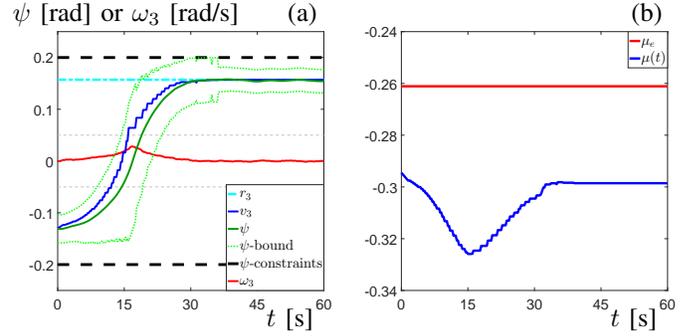

Fig.~\ref{fig:Ex_2} shows the simulation results, where the profiles of $(v_3, \psi, \omega_3)$ are presented in Fig.~\ref{fig:Ex_2}(a) to illustrate the system response and $v_{1,2}$, $\phi$, $\theta$, and $\omega_{1,2}$ respond in similar ways. The profiles of $(\psi, \omega_3)$ show that the system operates in the range described above. Fig.~\ref{fig:Ex_2}(b) shows the profile of $\mu(t) = \mu\big(f_x\big(x_v(v(t)),v(t)\big)\big)$ versus $\mu_e = -0.261$. It can be seen that $\mu\big(f_x\big(x_v(v(t)),v(t)\big)\big)$ along the trajectory of $x_v(v(t))$ is bounded by $-0.261$ from above, which serves as an example to show the fact that {\it (A6)} holds and $\mu_e = -0.261$ is an upper bound for logarithmic norms involved in our RG operation.

%%%%%%%%%%%%%%%%%%%%%%%%%%%%%%%%%%%%%%%%%%%%%%%%%%%%%%%%%%%%%
%%%%%%%%%%%%%%%%%%%%%%%%%%%%%%%%%%%%%%%%%%%%%%%%%%%%%%%%%%%%%

\section{Conclusion}\label{sec:7}

This note described a reference governor (RG) design for a continuous-time nonlinear system with an additive disturbance. The design is based on bounding (covering) the response of the nonlinear system by the response of a linear model with a set-bounded error, where the error bound is explicitly characterized using logarithmic norms. The online optimization is reduced to a convex quadratic program with linear inequality constraints, which can be easily solved. The proposed RG design guarantees sample-time and inter-sample constraint enforcement, recursive feasibility, and finite-time convergence of reference to command under appropriate assumptions.

% Theoretical properties including constraint enforcement, recursive feasibility, and finite-time convergence were analyzed.

\bibliographystyle{IEEEtran}
\bibliography{ref}

\end{document}